%% file: paper.tex
\documentclass[10pt,sigplan,letterpaper,twocolumn]{article}
\usepackage[10pt,nocopyright]{sigmin}

\usepackage{listings}
\usepackage{color}
\usepackage{makecell}
\usepackage[hyphens]{url}
\usepackage[breaklinks,colorlinks]{hyperref}
\usepackage[usenames,dvipsnames]{xcolor}
\hypersetup{citecolor=blue,linkcolor=blue}
\usepackage{amsmath,amsopn,amssymb,amsthm}
\usepackage{thmtools}
\usepackage[toc,page]{appendix}
\usepackage{subcaption}
\usepackage{endnotes,microtype,xspace,fancyvrb,multirow}
\microtypesetup{protrusion=false}
\usepackage[normalem]{ulem}
\usepackage{booktabs}
\usepackage{tabularx}
\usepackage{array,underscore,relsize}
\usepackage[T1]{fontenc}
\usepackage{times}
\usepackage{fancyhdr,lastpage}
\usepackage{enumitem}
\usepackage[labelfont=bf,font=normal,skip=1pt]{caption}
\usepackage[export]{adjustbox}
\usepackage{breakurl}
\usepackage{pifont}
\usepackage{tablefootnote}
\usepackage{bbding}
\usepackage[linesnumbered,vlined,boxed,commentsnumbered]{algorithm2e}
\DontPrintSemicolon
\usepackage{algpseudocode}
\usepackage{amssymb}
\usepackage[hang,flushmargin]{footmisc}
\usepackage{textcomp}
\usepackage{graphicx}
\usepackage{authblk}

\usepackage{colortbl}
\definecolor{lightgray}{RGB}{236,236,236}
\definecolor{redcolor}{RGB}{234,51,35}
\definecolor{greencolor}{RGB}{80,177,51}
\definecolor{blackcolor}{RGB}{0,0,0}

\usepackage[compact]{titlesec}
\titleformat*{\section}{\large\bfseries}
\titleformat*{\subsection}{\normalsize\bfseries}
\titleformat*{\subsubsection}{\normalsize\bfseries}
\titlespacing{\section}{0pt}{3ex}{1ex}
\titlespacing{\subsection}{0pt}{2ex}{1ex}

\makeatletter
\lst@AddToHook{OnEmptyLine}{\global\advance\c@lstnumber\m@ne}
\makeatother

\hypersetup{
  colorlinks,
  linkcolor={red!50!black},
  citecolor={blue!50!black},
  urlcolor={blue!80!black}
}

\newcommand{\sys}{\textsc{SMetric}}

\newcommand{\kvcache}{KV\$\xspace}
\newcommand{\nospacestitle}[1]{\noindent{\bf #1}}

\newcommand{\fig}[1]{Figure{~\ref{#1}}}

\newcommand{\stitle}[1]{\vspace{1.1ex}\noindent{\bf #1}}
\newcommand{\etitle}[1]{\vspace{1ex}\noindent{\em \ul{#1}}}

\newcommand{\company}{\textsc{Bailian}}

\usepackage{tikz}
\usetikzlibrary{shadows.blur}

\newcounter{finding}
\newcommand{\finding}[1]{%
  \par\vspace{1.2ex}\noindent
  \refstepcounter{finding}%
  \begin{tikzpicture}
    \node[draw=black!45, line width=0.4pt, fill=lightgray,
          rounded corners=2pt, inner sep=6pt, align=justify,
          blur shadow={shadow blur steps=8, shadow xshift=1pt,
                       shadow yshift=-1.5pt, shadow opacity=30},
          text width=\dimexpr\columnwidth-16pt\relax]
      {\textbf{Finding \thefinding:}
       #1};
  \end{tikzpicture}%
  \par\vspace{0.8ex}}

\newcommand{\takeaway}[1]{%
  \par\vspace{1.2ex}\noindent
  \begin{tikzpicture}
    \node[draw=black!45, line width=0.4pt, fill=lightgray,
          rounded corners=2pt, inner sep=6pt, align=justify,
          blur shadow={shadow blur steps=8, shadow xshift=1pt,
                       shadow yshift=-1.5pt, shadow opacity=30},
          text width=\dimexpr\columnwidth-16pt\relax]
      {\textbf{Takeaway:}
       #1};
  \end{tikzpicture}%
  \par\vspace{0.8ex}}

\usepackage{balance}
\usepackage{soul}

\begin{document}
\RestyleAlgo{ruled}

\title{\LARGE \bf{{\sys}: Rethink LLM Scheduling for Serving Agents with \\ Balanced Session-centric Scheduling}}

\setlength{\affilsep}{0.3em}
\renewcommand\Authsep{,\quad}
\renewcommand\Authand{,\quad}
\renewcommand\Authands{,\quad}
\makeatletter
\renewcommand\AB@affilsepx{ \quad\protect\Affilfont \, }
\makeatother
\author[1]{Jiahao Wang}
\author[1]{Kaizhan Lin\textsuperscript{\dag}}
\author[1]{Kaixi Zhang}
\author[1]{Jinbo Han}
\author[1]{Xingda Wei\,{\Envelope}}
\author[2]{Sijie Shen}
\author[2]{Chenguang Fang}
\author[2]{Wenyuan Yu}
\author[1]{Rong Chen}
\author[1]{Haibo Chen}

\affil[1]{\vspace{-2.mm}Institute of Parallel and Distributed Systems, Shanghai Jiao Tong University}
\affil[2]{Alibaba Group\vspace{-1.mm}}
\date{}

\maketitle
\begin{NoHyper}
\def\thefootnote{\dag}\footnotetext{Kaizhan Lin is affiliated with ShanghaiTech University; this work was done while he was an intern at Institute of Parallel and Distributed Systems, Shanghai Jiao Tong University.}
\def\thefootnote{\Envelope}\footnotetext{Xingda Wei is the corresponding author (\url{wxdwfc@sjtu.edu.cn}).}
\end{NoHyper}
\renewcommand{\thefootnote}{\arabic{footnote}}

\input{abs}

\input{body/intro.tex}

\input{body/bg-v1.tex}

\input{body/analyze-hit.tex}

\input{body/design.tex}

\input{body/eval.tex}

\input{body/related.tex}

\input{body/concl.tex}

\balance

\small{
\bibliographystyle{plain}
\bibliography{paper}
}

\input{body/appendix.tex}

\end{document}

%% file: abs.tex
\begin{abstract}

\noindent
LLM scheduling is critical to serving,
yet how well existing designs fit agentic
serving---where agents, not humans, issue the
requests---remains unclear. Agents shift the workload in two
ways: they consume many more tokens than humans, so the cluster
must provide high throughput (tokens per second, TPS) at low latency;
and their requests reuse far
more {\kvcache} than chat.

Existing schedulers still trade off load balance against {\kvcache}
reuse under the new workloads: cache-aware schedulers may crowd requests onto the few instances
caching the {\kvcache}, leaving the rest idle, while balanced
schedulers may lose the opportunity for reuse,
which is costly, as the benefit of reuse becomes more dominant
at a high reuse ratio.
We thus present two key insights: (1) with a global-tier {\kvcache}
store, pursuing load balance need not compromise {\kvcache}
reuse, though the slower global tier must be used with
care;
and (2) given the agent's intra-session locality, routing
requests by their sessions can balance the load with high
{\kvcache} reuse.

A key challenge in realizing session-centric scheduling is that
the scheduler must identify a request's session statelessly,
which is difficult for model providers serving arbitrary agents.
{\sys} addresses this with differential scheduling based on two
indicators derived from the request itself, the \emph{session turn} and
the local {\kvcache} hit: it schedules first-turn requests for
load balance, as they mark the arrival of new sessions, and sticks
follow-ups to the instance with the highest local hit for high {\kvcache} reuse.
As sessions differ widely in size, {\sys} sticks a follow-up only
if the instance can serve it within its SLO, and
otherwise migrates the session to the least-loaded instance
to prevent many long sessions from eventually imbalancing the load.
Evaluated on real-world traces, {\sys}
improves the peak TPS by 9--15\,\% under prefill-decode
colocation with a provisioned global tier and the peak prefill TPS by
9\,\% under disaggregation over state-of-the-art schedulers, also with
lower latency.

\end{abstract}

%% file: body/intro.tex
\section{Introduction}
\label{sec:intro}

\noindent
Large language models (LLMs) now power agents like Claude Code~\cite{claudecode,openclaw}
that accomplish complex tasks
such as
writing~\cite{anthropic2025ccompiler}, debugging~\cite{ding2026fmagent}, and performance-tuning~\cite{qwen3.7,minimax-m3} code.
An agent solves a task through a \emph{session} of LLM requests,
each of which requires generating many tokens.
To generate these tokens efficiently,
providers deploy clusters dedicated to serving agent requests;
each cluster runs many instances, each holding a full copy of the model.
We term this cluster-scale, agent-dedicated LLM serving \emph{agentic serving}.
Agentic serving is still LLM serving:
the provider serves each request through the standard LLM API
and neither runs the agents nor sees their internal logic.

Request scheduling---how the cluster routes each request to an instance---is a key
pillar of LLM serving.
Though recent work studies it extensively~\cite{zhang2026simple,DBLP:journals/corr/abs-2507-17769,DBLP:conf/iclr/SrivatsaHAL025,DBLP:journals/corr/abs-2602-06502,llm-d,aibrix},
we argue that scheduling designs need to be revisited for agents,
for two reasons:

First,
an agent consumes far more tokens
per request than a human user~\cite{bai2026aiagentsspendmoney},
so a cluster must
provide high throughput (tokens per second, TPS) to serve the load at low cost.
Besides cost, the latency---the time to first token
(TTFT) and the time per output token (TPOT)~\cite{andes}---cannot be compromised either.
This is because the token generation latency determines how fast agents finish their tasks,
and a response that is too slow makes the agent treat the request as
failed~\cite{claude-opus4-outage}.

Second, agentic serving exhibits a different workload pattern from chat:
{\kvcache} reuse dominates.
The efficiency of serving an LLM request depends heavily on {\kvcache} reuse;
the {\kvcache} is the intermediate state a cluster caches to avoid
recomputation (\textsection{\ref{sec:bg}}).
In chat workloads, this reuse is modest: only 54--62\,\% of request tokens are reused from
cache~\cite{kvcache}.
In agentic serving, it is far higher:
our analysis of the production trace from a cluster at a
representative model provider (\company) reveals that
the reuse exceeds 80\,\% (Finding~\ref{finding:reuse} in \textsection{\ref{sec:trace-analyze}}).
This shift requires revisiting how well scheduling designs facilitate
{\kvcache} reuse, because the computation saved by reuse becomes more
dominant at a high reuse ratio: raising it from 70\,\% to 80\,\% cuts
the prompt tokens still to compute by one third.

In this paper, we first conduct a systematic study of request scheduling for
agents, using two recent large-scale real-world agentic serving traces (\textsection{\ref{sec:trace-analyze}}).
We find that state-of-the-art schedulers still suffer from the
trade-off between
{\kvcache} reuse and load balance~\cite{zhang2026simple}.
Specifically, to achieve high reuse, the current scheduler in {\company}
prioritizes routing each request to the
instance holding its cached {\kvcache} (cache-aware).
This trades away load balance: some instances overload while others
sit idle, capping the cluster's throughput.
On the other hand, a balanced scheduler like LMetric~\cite{zhang2026simple}
trades away {\kvcache} reuse, requiring more computation to serve the requests,
so the throughput gain from better load balancing is limited.

To address this dilemma, we draw two insights.
First, from the system's perspective, pursuing load balance does not have to
compromise {\kvcache} reuse:
modern agentic serving adopts a two-tier {\kvcache} store---a local tier on
each instance's GPU memory and a global tier on cluster-wide CPU memory
(\textsection{\ref{sec:bg}}).
Even if the scheduled instance does not hold the {\kvcache} locally, it can
fetch it from the global tier.
As long as the global tier serves these {\kvcache} fetches in time, TPS
remains high (\textsection{\ref{sec:analysis-methods}}).
The global tier, however, must be used with care:
it is slower---waiting for the {\kvcache} to be transferred adds
latency---and its capacity and bandwidth are limited.
Routing \emph{every} request purely for balance---even
with global-tier {\kvcache} reuse---is thus still suboptimal.

Second, from the agent workload's perspective, routing requests by
their \emph{sessions} can balance the load with high {\kvcache} reuse.
On the one hand, sessions are a natural unit for balancing: though
individual sessions have skewed token usage, the token load
\emph{can} still be balanced across instances at scale
(Finding~\ref{finding:balance}).
On the other hand, sessions capture most of the reuse, thanks to the
workload's intra-session locality: follow-up requests mostly reuse
the {\kvcache} of earlier turns in the same session
(Finding~\ref{finding:intra}).
Keeping each session's requests on one instance thus preserves
most of the {\kvcache} reuse in its local tier, while spreading sessions
across instances balances the cluster.
At first glance, the {\kvcache} reuse of first-turn requests seems
compromised.
The impact is small: their reuse comes mostly from the system prompt
(Finding~\ref{finding:system-prompt}), which the global tier serves
efficiently.

One key challenge in realizing such session-centric scheduling is that
the scheduler must identify each request's session \emph{statelessly},
which is difficult for model providers serving arbitrary agents:
the request carries no session information, as the LLM APIs are
stateless, and its source may further be anonymized by proxies.
Inferring sessions from records of past requests is typically impractical, as
such per-session state grows with active sessions,
making a simple and scalable router design challenging.

We address this challenge with {\sys}.
The key observation is that the \emph{session turn} and the local
{\kvcache} hit, both derived from the request itself, identify
sessions statelessly and at zero cost.
The turn, deduced from the number of historical messages in the request,
tells the scheduler whether the request starts a session:
{\sys} thus schedules first-turn requests for load balance,
spreading sessions across the cluster (\textsection{\ref{sec:design}}).
Meanwhile, the local {\kvcache} hit, which the router already knows,
helps identify the instance serving the session,
because
intra-session reuse happens quickly (Finding~\ref{finding:speed}),
so the instance with the highest hit is likely the one that executed
the session's previous request.
Therefore, {\sys} sticks follow-up requests to it.
This differential scheduling achieves load balance and high
local-tier reuse while keeping demand on the global tier low, because
in the ideal case only first-turn requests fetch {\kvcache} from it.

While such differential scheduling identifies sessions statelessly,
we meet a second challenge: sticking every follow-up request to its
session's instance still imbalances the load, because
sessions differ widely in size (\textsection{\ref{sec:analyze-session}})
and the size is unknown when the session is first routed,
so an instance overloads as multiple long sessions
are scheduled onto it.
{\sys} thus sticks a follow-up request only if the instance can finish
it within its TTFT SLO, and otherwise migrates the session by routing
the request to the least-loaded instance, with the global tier
preserving its {\kvcache} reuse (\textsection{\ref{sec:design}}).

We have implemented {\sys} on vLLM~\cite{vllm-code} and LMCache~\cite{lmcache},
and extensively evaluated it against state-of-the-art
schedulers, including both the production scheduler used in {\company}
and recent research and open-source schedulers
(LMetric~\cite{zhang2026simple}, llm-d~\cite{llm-d},
Preble~\cite{DBLP:conf/iclr/SrivatsaHAL025}, and Dynamo~\cite{ai-Dynamo}),
with serving paradigms spanning both prefill-decode colocation and prefill-decode disaggregation.
Against the best-performing baseline,
{\sys} serves 9--15\,\% more peak TPS under colocation
with a provisioned global tier,
and 9\,\% more peak prefill TPS under disaggregation;
the gap widens to 26--28\,\% as the load grows.
It also achieves the lowest median TTFT and
the lowest P90 TPOT under colocation,
and the lowest tail TTFT under disaggregation
(19\,\% lower at P90 than the closest baseline).

In summary, this paper makes the following contributions:
\begin{itemize}[leftmargin=*,leftmargin=10pt,itemindent=0pt]
    \item The first study of the workload patterns of LLM requests served by
    an agent-dedicated cluster at scale (\textsection{\ref{sec:trace-analyze}}).
    \item The first analysis of how existing schedulers perform with the
    two-tier {\kvcache} store under agentic serving (\textsection{\ref{sec:analysis-methods}}).
    \item A session-centric scheduling design that identifies sessions
    statelessly by the turn and the local {\kvcache} hit, and migrates
    overloaded sessions by their SLO, achieving both high {\kvcache} reuse
    and load-balanced serving (\textsection{\ref{sec:solution-v2}}).
    \item An evaluation showing that {\sys} improves TPS and latency over
    state-of-the-art schedulers (\textsection{\ref{sec:eval}}).
\end{itemize}

We will open-source {\sys} and the traces upon publication.

%% file: body/bg-v1.tex
\section{Background and System Setup}
\label{sec:bg}

\begin{figure}[!t]
        \vspace{2mm}
        \begin{minipage}{1\linewidth}
        \centering
        \includegraphics[width=0.99\linewidth]{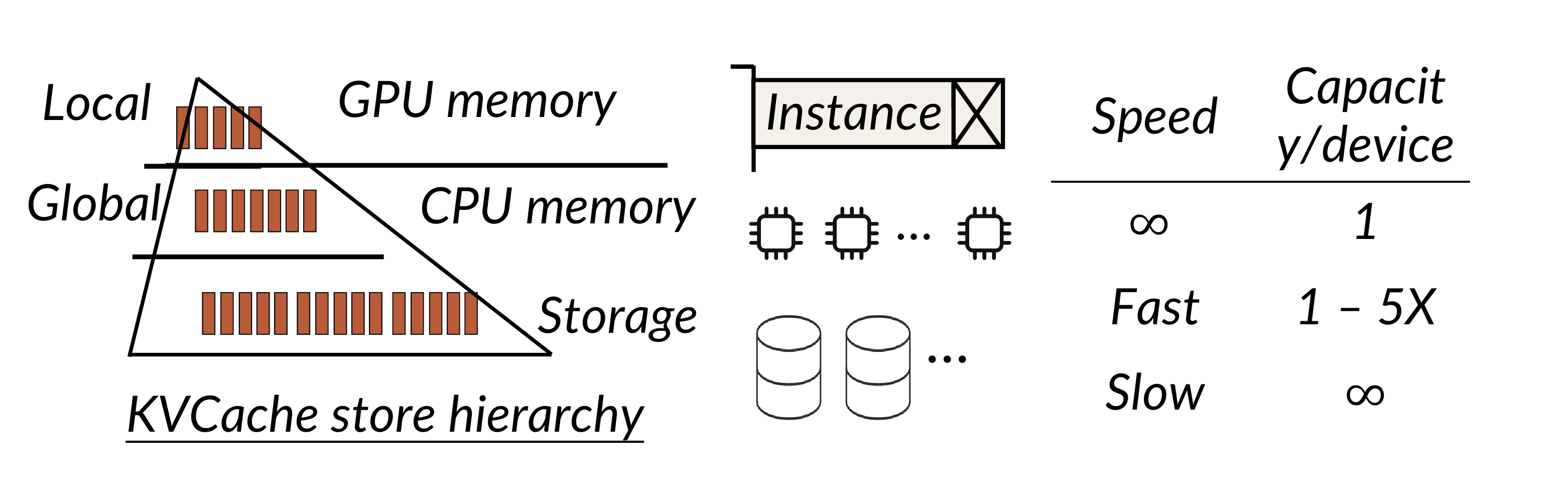}
        \end{minipage} \\[4pt]
        \begin{minipage}{1\linewidth}
        \caption{\small{%
            The {\kvcache} store hierarchy for serving agents.
        }}
        \label{fig:kvcache}
        \end{minipage} \\[-15pt]
        \end{figure}

\nospacestitle{LLM and {\kvcache}. \,}
An LLM serves a request in two phases.
Given the input tokens (the \emph{prompt}),
a \emph{prefill} pass processes the prompt with a single model forward (possibly split into chunks~\cite{298679})
and emits the first output token.
A \emph{decode} pass then emits the remaining tokens auto-regressively:
each token comes from one model forward over the prompt plus all tokens emitted so far,
until the model produces an end-of-sequence (EOS) token.

The forward pass of LLMs requires computing a key (K) and a value (V)
tensor for each token in the input sequence.
Since these tensors are the same for inputs with the same prefix,
it is standard to cache the KV tensors generated so far (the {\kvcache}) to accelerate the decode phase,
as well as the prefill phase of later requests with the same prefix.
As shown in {\fig{fig:pdd}}~(a),
if a 3-token input request has two {\kvcache} hits,
then it needs to compute the {\kvcache} for only one token.

\stitle{{\kvcache} store hierarchy and the local vs. global tiers. \,}
Since cross-request reuse is common in production workloads~\cite{kvcache,305212},
serving systems typically cache the {\kvcache} of finished requests for some time.
Caching requires storage capacity because each request's {\kvcache} is large:
for example, a single 128K-token request consumes about 32\,GB of {\kvcache} on the popular Qwen3-32B model.
As a result, the {\kvcache} store in production systems follows a multi-tier architecture
as shown in \fig{fig:kvcache}.
The first tier is device (GPU) memory;
when it fills, {\kvcache} cascades to the larger, global tiers below---host (CPU) memory,
then cloud storage (e.g., S3).
A later {\kvcache} hit on a lower tier must load the entry back up to the GPU.

\begin{figure}[!t]
        \vspace{2mm}
        \begin{minipage}{1\linewidth}
        \centering
        \includegraphics[width=0.94\linewidth]{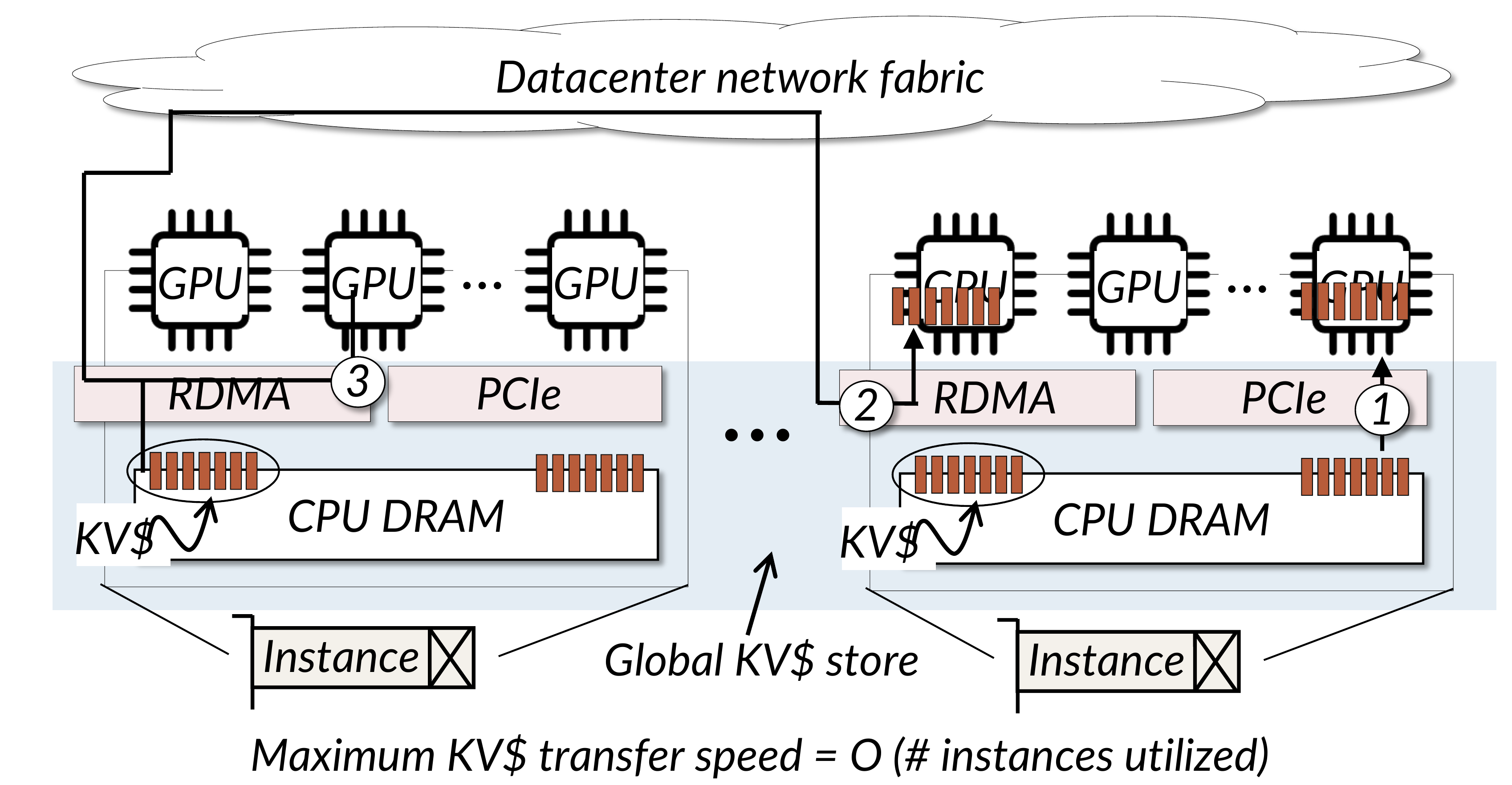}
        \end{minipage} \\[7pt]
        \begin{minipage}{1\linewidth}
        \caption{\small{%
        The detailed architecture of global-tier {\kvcache}.
        The global tier enables a GPU to read {\kvcache} from local CPU (\ding{192}),
        remote CPU memory (\ding{193}) and remote GPU memory via GPU-direct RDMA (\ding{194}).
        }}
        \label{fig:global-tier}
        \end{minipage} \\[-10pt]
\end{figure}

\begin{figure}[!t]
        \vspace{2mm}
        \begin{minipage}{1\linewidth}
        \centering
        \includegraphics[width=0.99\linewidth]{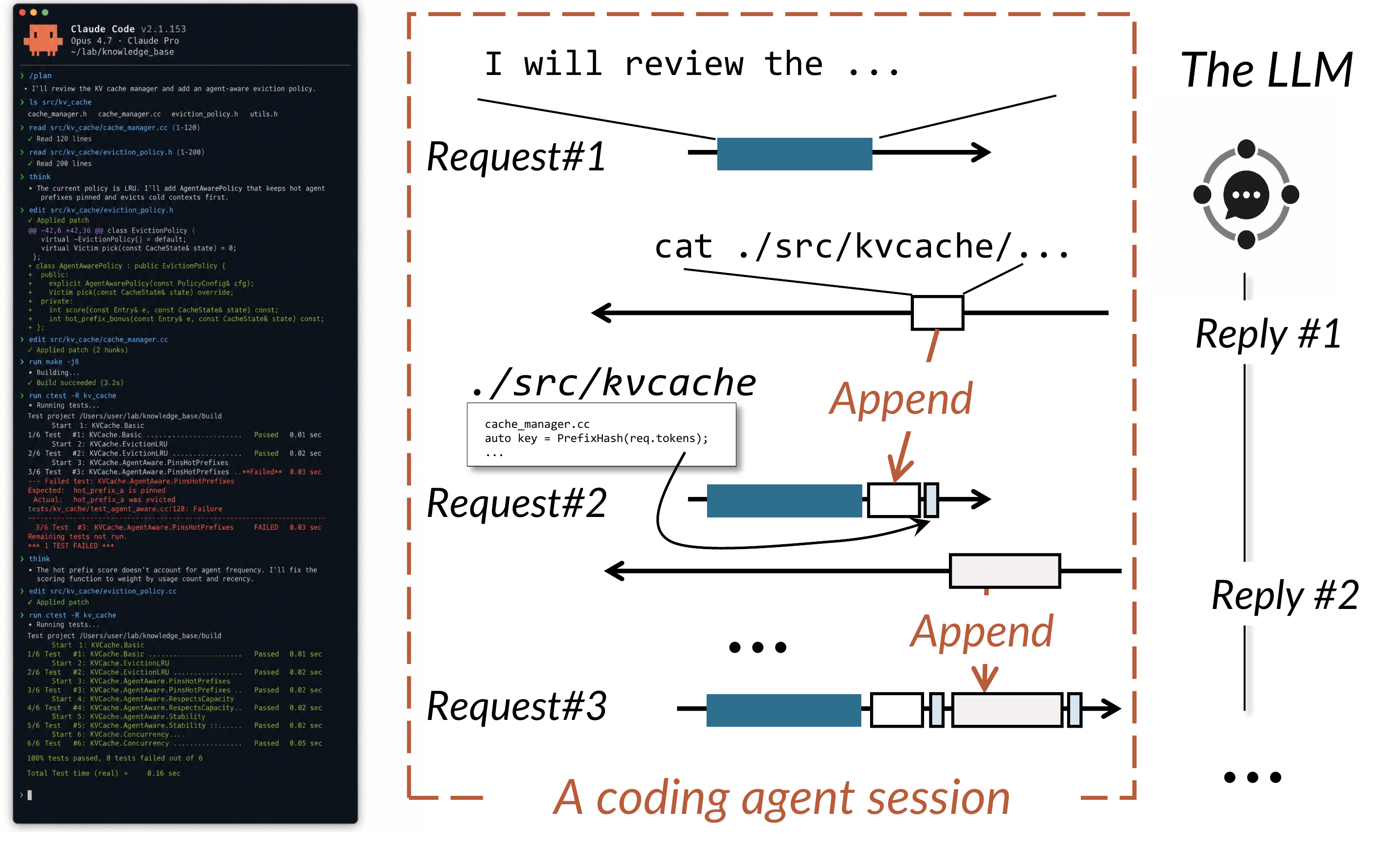}
        \end{minipage} \\[7pt]
        \begin{minipage}{1\linewidth}
        \caption{\small{%
            An example of how agents interact with the served LLM.
        }}
        \label{fig:agent}
        \end{minipage} \\[-15pt]
        \end{figure}

\stitle{Agentic workflow and the agent session. \,}
Modern LLM agents such as Claude Code~\cite{claudecode} solve a task through a \emph{session} of
multiple rounds of interaction with the LLM.
As shown in \fig{fig:agent},
the first request (Request~\#1) in a session, typically written by the user, specifies the task to solve.
The LLM replies with tokens that carry its reasoning and the next step for the agent,
such as a tool call that reads a file with \texttt{cat} (Reply~\#1).
The agent then runs the tool, appends the tool call output to the context, and issues the next request (Request~\#2).
Because the LLM keeps no state across requests, each request resends the whole history.
This append-only pattern---request~$i$ extends request~$i{-}1$---lets request~$i$ hit the {\kvcache} of request~$i{-}1$,
if the platform has not evicted it.

\stitle{LLM serving architecture in the wild. \,}
To serve LLM requests at scale, providers typically deploy each model on a cluster of GPUs\footnote{\footnotesize{Providers may deploy multiple clusters; inter-cluster scheduling is beyond the scope of this work.}}.
The smallest unit of deployment is a \emph{model instance}---one or more GPUs that together hold a full copy of the model parameters.
A \emph{global router} dispatches each incoming request to one instance using a routing algorithm~\cite{DBLP:conf/osdi/SunHZXZL024,DBLP:conf/iclr/SrivatsaHAL025,DBLP:journals/corr/abs-2507-17769,zhang2026simple,skywalker}.
Because a request runs in two phases---prefill and decode---there are two mainstream serving architectures today:

\begin{figure}[!t]
        \vspace{2mm}
        \begin{minipage}{1\linewidth}
        \centering
        \includegraphics[width=0.95\linewidth]{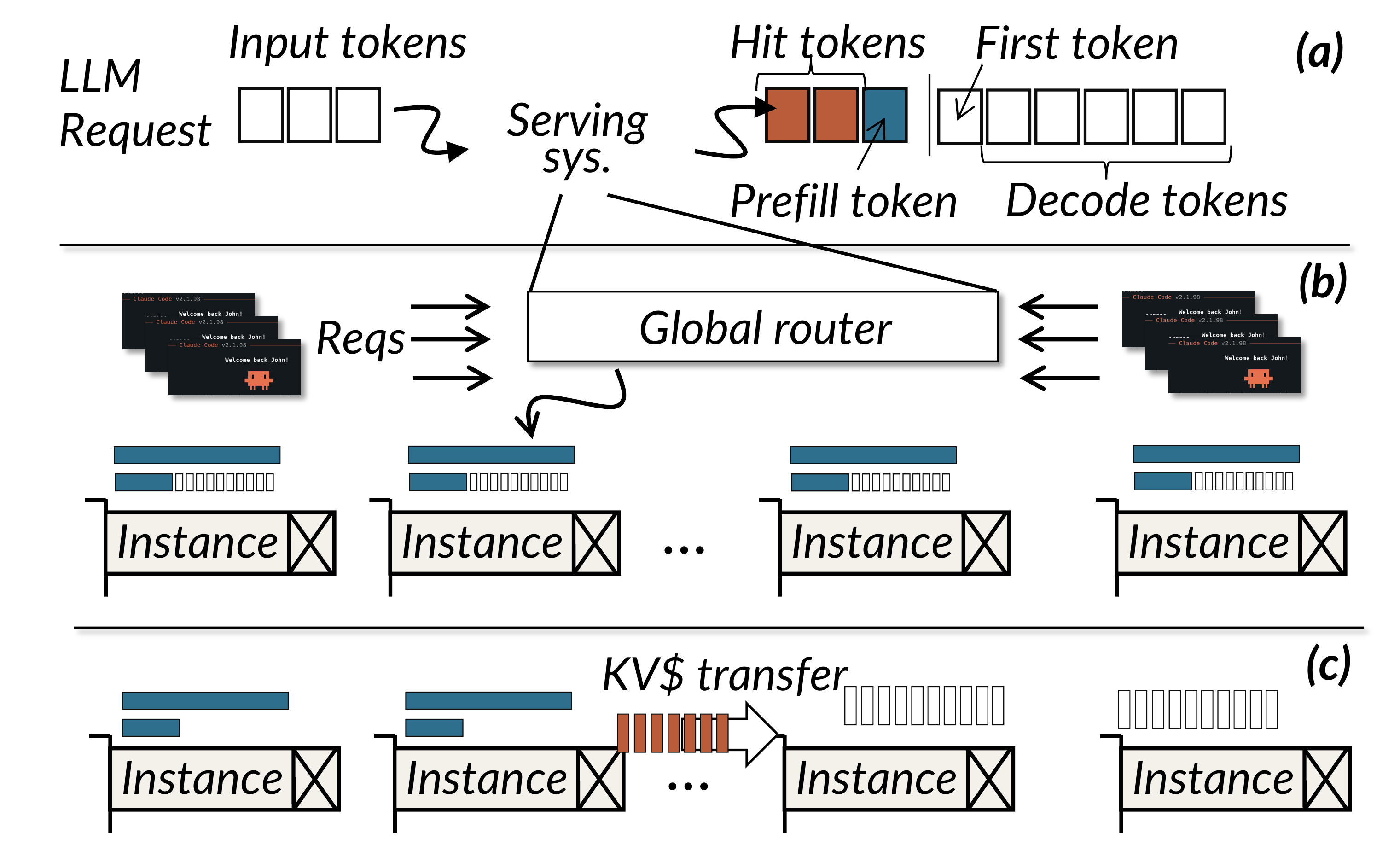}
        \end{minipage} \\[7pt]
        \begin{minipage}{1\linewidth}
        \caption{\small{%
            (a) An overview of the prefill and decode of LLM processing with {\kvcache}.
            (b) The PD-colocated and
            (c) the PD-disaggregated serving architecture.
        }}
        \label{fig:pdd}
        \end{minipage} \\[-0pt]
        \end{figure}

\etitle{PD colocation. \,}
PD colocation runs the prefill and decode of a request to completion on the same instance ({\fig{fig:pdd}}~(b)).

\etitle{PD disaggregation. \,}
PD disaggregation uses two clusters for serving:
one cluster only runs prefill
and the other only runs decode~\cite{DBLP:conf/osdi/ZhongLCHZL0024,DBLP:conf/isca/PatelCZSGMB24} ({\fig{fig:pdd}}~(c)).
A request first runs prefill on a prefill instance,
which transfers the resulting {\kvcache} over RDMA to a decode instance,
where the decode phase runs to completion.

Both architectures build atop the same {\kvcache} hierarchy of \fig{fig:kvcache}
and differ only in which instances access the global tiers.
Since cross-request {\kvcache} reuse happens at prefill,
every instance accesses the global tiers under PD colocation,
while only prefill instances do under PD disaggregation.

%% file: body/analyze-hit.tex
\section{Agentic serving patterns in the wild}
\label{sec:trace-analyze}

\nospacestitle{Analyzed traces. \,}
Existing production workload studies characterize LLM serving in general~\cite{servegen,305212}.
To understand the workload characteristics of real-world agentic serving,
we collected two recent representative traces from two large-scale clusters,
both dedicated to serving agents for a commercial coding-agent product (a paid subscription plan).
Each cluster is one of many supporting the product and contains at least thousands of accelerators.
Both traces cover the same peak usage hours (10:00--12:00) on May 29, 2026:
one cluster serves a popular TB-level model (Trace 1) and the other a roughly 280B-parameter model (Trace 2),
covering agents with different capabilities.

To study the workload in depth and to evaluate the serving system under it,
we collected as much information as possible within {\company}'s privacy policy.
For each LLM request, each trace records
(1) the hashed request content, which lets us analyze how much the request can benefit from {\kvcache} reuse,
(2) the arrival time, which captures request burstiness and {\kvcache} reuse intervals,
and (3) the session information---the session ID, the parent request ID, the turn number,
and how the request is triggered (e.g., by a human or by the agent).
Since requests carry no session identifier,
we reconstruct the session ID and the parent request ID offline during trace collection
(\textsection{\ref{sec:appendix-session}}).

\begin{figure}[!t]
    \vspace{2mm}
    \begin{minipage}{1\linewidth}
    \centering
    \includegraphics[width=.95\linewidth]{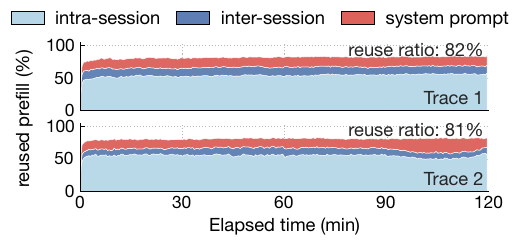}
    \end{minipage} \\[5pt]
    \begin{minipage}{1\linewidth}
    \caption{\small{%
        A characterization of the reusable {\kvcache} for all requests,
        with a breakdown of where the reuses come from.
    }}
    \label{fig:kvcache-reuse-breakdown}
    \end{minipage} \\[-15pt]
\end{figure}

\subsection{{\kvcache} reuse patterns}
\label{sec:analyze-kvc}

\noindent
We analyze three metrics of {\kvcache} usage in agentic serving---the reuse ratio,
where the reuse comes from, and the reuse interval---and distill four findings:

\finding{{\kvcache} reuse is \uline{dominant}:
the reuse ratio exceeds 80\,\% of the {\kvcache} in both traces ({\fig{fig:kvcache-reuse-breakdown}}).
\label{finding:reuse}
 }

We measure reuse at {\kvcache} block granularity against an idealized {\kvcache} store with infinite capacity,
using the block size of the traced cluster.
A request's reusable ratio is the fraction of its prefill tokens that fall in hit blocks,
i.e., the tokens whose prefill computation can be skipped.

\finding{{\kvcache} reuse follows \uline{session locality}: more than 65\,\% of reuses come from \emph{follow-up} requests (the second turn onward) reusing earlier turns of the same session ({\fig{fig:kvcache-reuse-breakdown}}).
\label{finding:intra}
}

For each reuse, we track which session originally produced the reused tokens.
{\fig{fig:kvcache-reuse-breakdown}} further breaks down the reuse on this basis into
intra-session reuse, when a request's reused part comes from a previous request in the same session,
and inter-session reuse, when it comes from a request in a different session.
Intra-session reuse contributes most of the reuse (roughly 67\,\% in both traces),
which is unsurprising: different sessions tend to solve different problems,
so cross-session reuse is rare.
Much of the remaining inter-session reuse stems from multi-session workflows~\cite{claudecodeagents2024,kimi2026agentswarm},
in which an agent forks multiple sub-agents whose sessions share the parent session's context.
Such attribution by the original producer also makes 67\,\% a lower bound on session locality:
when a sub-agent's follow-up request reuses its own earlier turns,
we count the reuse as inter-session, though it could also be counted as intra-session.

\finding{\uline{First-turn requests enjoy {\kvcache} reuse}: the shared system prompt accounts for 18--20\,\% of all reuses ({\fig{fig:kvcache-reuse-breakdown}}).
\label{finding:system-prompt}
}

\begin{figure}[!t]
    \vspace{2mm}
    \begin{minipage}{1\linewidth}
    \centering
    \makebox[\linewidth][c]{\includegraphics[width=0.95\linewidth]{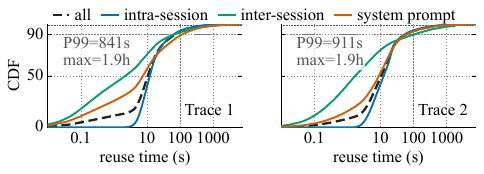}}
    \end{minipage} \\[3pt]
    \begin{minipage}{1\linewidth}
    \caption{\small{%
        Reuse-time distribution across the two production traces.
        The tail is truncated by the 2-hour trace collection window.
    }}
    \label{fig:kvcache-reuse-time-cdf}
    \end{minipage} \\[-5pt]
\end{figure}

A first-turn request has no earlier turn in its session to reuse,
yet it still hits on the system prompt:
requests from the same agent framework, e.g., Claude Code~\cite{claudecode} or OpenClaw~\cite{openclaw},
share one prompt template for the first request of each session.
This is also a case of inter-session reuse.

\begin{figure}[!t]
    \vspace{2mm}
    \begin{minipage}{1\linewidth}
    \centering
    \includegraphics[width=.99\linewidth]{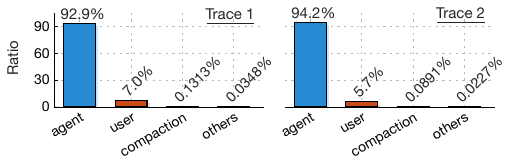}
    \end{minipage} \\[2pt]
    \begin{minipage}{1\linewidth}
    \caption{\small{%
    A breakdown of what triggers the requests in the two traces.
    \emph{Agent} denotes requests automatically sent by the agent; \emph{user} denotes requests sent by the human;
    \emph{compaction} is a special agent-triggered request that summarizes the context of the current session.
    }}
    \label{fig:req-trigger-breakdown}
    \end{minipage} \\[-10pt]
\end{figure}

\finding{{\kvcache} is \uline{quickly reused}:
about 90\,\% of all reuses recur within roughly 100\,s after the last request that shares the {\kvcache},
with only a small tail ({\fig{fig:kvcache-reuse-time-cdf}}).
\label{finding:speed}
}

{\fig{fig:kvcache-reuse-time-cdf}} plots the reuse interval---the time
from when a request completes until its {\kvcache} is next reused.
Reuse is quick, often within seconds,
because both traces are dominated by agent-triggered requests ({\fig{fig:req-trigger-breakdown}}):
the agent issues the next request as soon as it sees the previous response and tool results.
A few outliers are reused hours later,
likely when a user leaves the agent waiting for a response.

\subsection{Session execution patterns}
\label{sec:analyze-session}

\noindent
Next we examine the detailed pattern of each session
and derive the following two key findings:

\begin{figure}[!t]
    \vspace{2mm}
    \begin{minipage}{1\linewidth}
    \centering
    \includegraphics[width=.99\linewidth]{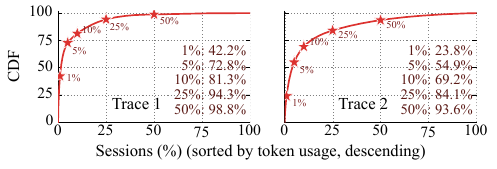}
    \end{minipage} \\[2pt]
    \begin{minipage}{1\linewidth}
    \caption{\small{%
        An analysis of token usage of different sessions.
    }}
    \label{fig:workload-session-skewness}
    \end{minipage} \\[-10pt]
\end{figure}

\finding{Session token usage is \uline{skewed}:
in both traces, the top 25\,\% of sessions contribute more than 80\,\% of the
tokens ({\fig{fig:workload-session-skewness}}).}

As shown in {\fig{fig:workload-session-skewness}},
the computation required varies significantly across sessions,
with a small fraction of sessions accounting for most of the tokens to compute.
Nevertheless, thanks to the large request volume,
the skewed computation can still be balanced across all instances:

 \finding{Despite the skewness, the token load
 can still be \uline{balanced} across serving instances over time. \label{finding:balance}}

\begin{figure}[!t]
    \vspace{2mm}
    \begin{minipage}{1\linewidth}
    \centering
    \includegraphics[width=.95\linewidth]{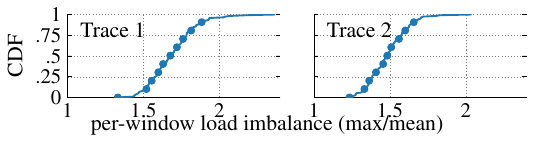}
    \end{minipage} \\[2pt]
    \begin{minipage}{1\linewidth}
    \caption{\small{%
        Per-window token load imbalance (max/mean) across instances
        under load-balancing dispatch, for two traces.
    }}
    \label{fig:sim-token-balance}
    \end{minipage} \\[-10pt]
\end{figure}

To quantify this, we use a trace-driven simulation that
replays all requests of a trace in arrival order
on a simulated cluster with the same number of instances as the traced cluster.
The simulator dispatches each request to the instance currently holding the fewest in-flight tokens,
counting all tokens---prefill and decode---of the requests running on that instance.
Each request stays at its instance for the execution time recorded in the trace.
{\fig{fig:sim-token-balance}} shows the resulting load balancing profile:
for each one-minute window, we report the token count of the most loaded instance
divided by the mean token count across all instances (max/mean).
The load stays largely balanced:
the median per-window imbalance is only 1.7 for Trace 1 and 1.5 for Trace 2,
and even the worst window stays below 2.4 and 2.1, respectively.
The simulation is approximate---a request's execution time on the simulated
cluster may differ from that on the original cluster---but it suffices to
reveal the workload's balance potential,
and our analysis in \textsection{\ref{sec:analysis-methods}} further confirms that
schedulers balancing the load alone keep instance loads balanced.

Balance holds because concurrent requests far outnumber instances:
the number of requests arriving within a 1-minute window is on average
43$\times$ (Trace 1) and 48$\times$ (Trace 2) the number of serving instances,
and never drops below 30$\times$ in either trace.
Hence, even when token-heavy requests occupy a subset of instances,
the remaining instances absorb many token-light ones,
which evens out the skew of individual sessions.
Such a high request-to-instance ratio is common in serving:
a GPU instance requires a large batch of requests to achieve high utilization,
and requests are not scarce for large service providers like {\company}.

%% file: body/design.tex
\section{Balanced session-centric scheduling}
\label{sec:design}

\subsection{Analysis of existing schedulers}
\label{sec:analysis-methods}

\lstdefinestyle{ssched}{language=Python,numbers=left,
    xleftmargin=1.9em,framexleftmargin=0pt,
    frame=single,framerule=0.2pt,framesep=2pt,rulecolor=\color{black!35},numbersep=5pt,
    backgroundcolor=\color{blue!4},
    numberstyle=\scriptsize\color{black!45},
    numberblanklines=false,
    basicstyle=\scriptsize\ttfamily,identifierstyle=,
    keywordstyle=\bfseries\color{black},keywordstyle=[2]{},
    commentstyle=\color{black!55},
    alsoletter={_},
    emph={c,hit,hit_len,est_hit,session_cached},
    emphstyle=\color[RGB]{0,110,110},
    emph=[2]{l,load,q,queued_cost,queued_prefill,estimate_load,prefill_cost,num_requests,C_LIN,C_ATT,PREFILL_RATE},
    emphstyle=[2]\color[RGB]{112,48,160},
    deletekeywords={max},deletekeywords=[2]{max},
    emph=[3]{argmin,max,argmax,rr_argmin,f,SLACK},
    emphstyle=[3]\bfseries\color{black},
    emph=[4]{HIT_RATIO},emphstyle=[4]\bfseries\color[RGB]{0,110,110},
    escapechar=@,aboveskip=2pt,belowskip=0pt,lineskip=1.5pt}
\newcommand{\lstregion}[1]{\rlap{\hbox to \linewidth{\hfill\textsf{\bfseries\scriptsize #1}}}}
\newcommand{\lstbar}{\textcolor{orange!85!black}{\rule[-0.35ex]{1.4pt}{1.6ex}}}
\newcommand{\lstchg}{\llap{\lstbar\kern3.2pt}}

\begin{figure}[t]
\begin{lstlisting}[style=ssched,firstnumber=1]
req = receive()
c = kvc.hit_len(req)   # local KV$ hit, per instance
l = load()             # load, per instance

sched_to = instances.argmin/max(f(c, l))
req.forward(sched_to)
\end{lstlisting}
\vspace{3pt}
\caption{\small{
    A high-level overview of the existing LLM scheduling scoring methods.
    $f(c_i, l_i)$ scores instance $i$ by the request's {\kvcache} reuse
    on it ($c_i$) and its load ($l_i$).
}}
\label{fig:general-scheduler}
\end{figure}

\nospacestitle{A preliminary on LLM scheduling in a cluster. \,}
To the best of our knowledge, existing LLM
schedulers~\cite{zhang2026simple,305212,llm-d,aibrix,ai-Dynamo,aigw,DBLP:conf/iclr/SrivatsaHAL025,DBLP:journals/corr/abs-2507-17769,DBLP:journals/corr/abs-2602-06502,DBLP:journals/corr/abs-2504-08784,skywalker}
route requests by \emph{per-request scoring}, as
{\fig{fig:general-scheduler}} illustrates:
upon receiving a request, the scheduler computes a score for each
instance and routes the request to the instance with the lowest
(or highest, depending on the definition) score.
The score of instance $i$ is typically a function $f(c_i, l_i)$ of two LLM-specific indicators:
$l_i$ describes the instance's current state, typically its load,
e.g., its batch size or the number of inflight tokens.
$c_i$ estimates the computation the request would need on the instance,
typically by the prefix hit length of the request's prompt in
instance $i$'s local {\kvcache}.
Methods mainly differ in how they define $f$,
i.e., how they weigh the instance's load against the request's
{\kvcache} reuse.
Our motivating analysis uses the following representative methods,
while \textsection{\ref{sec:eval-setup}} covers more: \\[-10pt]
\begin{itemize}[leftmargin=*,leftmargin=10pt,itemindent=0pt]
    \item \emph{{\company}} is the production scheduler deployed at {\company}.
        It scores an instance by a weighted sum of its load and the
        request's {\kvcache} miss:
        $f = (1-\lambda) \cdot \hat{l}_i + \lambda \cdot (1 - \hat{c}_i)$
        (the lower the better),
        where $\hat{l}_i$ is the instance's batch size and
        $\hat{c}_i$ is the request's {\kvcache} hit ratio on the instance.
        Each indicator carries a $\hat{\cdot}$ because it is normalized
        to a common range; otherwise the two cannot be added.
        The weight $\lambda$ is tuned to trade {\kvcache} reuse for load
        balance; for agentic workloads, the operator sets a large
        $\lambda$ to favor {\kvcache} reuse.
        Popular open-source schedulers adopt a similar
        design~\cite{ai-Dynamo,aigw}.
         \\[-10pt]

    \item \emph{LMetric}~\cite{zhang2026simple} is a recent scheduler that
        multiplies the two indicators instead of adding them for scoring instances:
        $f = (q_i + L - c_i) \times l_i$,
        where $L$ is the request's prompt length and
        $q_i$ is the prefill work already queued at instance $i$,
        so $q_i + L - c_i$ is the prefill computation left before the
        request's first token.
         \\[-10pt]

    \item \emph{Load-balance-only} considers load only, i.e., $f = l_i$:
        it routes each request to the least-loaded instance
        regardless of its {\kvcache} reuse.
        It is a variant of the classic join-the-shortest-queue policy
        used in traditional cluster-level request routing,
        and the default policy of vLLM~\cite{vllm-code}. \\[-10pt]
\end{itemize}

\stitle{The analysis methodology. \,}
To understand how representative methods perform when serving agents,
this section compares the methods by replaying Trace~1 on a 32-instance
vLLM~\cite{vllm-code} cluster serving Qwen3-Coder-30B-A3B under PD
colocation, with Mooncake~\cite{305212} and LMCache~\cite{lmcache}
as the global {\kvcache} store;
\textsection{\ref{sec:eval-setup}} details the testbed, its tuning,
and how the replay keeps the {\kvcache} reuse consistent with the trace.
We replay only Trace~1, since the two traces exhibit similar patterns
(\textsection{\ref{sec:trace-analyze}});
PD disaggregation shows similar results
(\textsection{\ref{sec:eval-e2e-pd}}).

We report each method's \emph{TPS within SLO}:
the cluster-wide output TPS (output tokens generated per second)
contributed only by requests that satisfy both the
TTFT and TPOT SLOs.
The request-level SLO matters for agents: it determines the task
completion time, and a response that is too slow makes the agent treat
the request as timed out~\cite{claude-opus4-outage}.
The SLO follows common practice~\cite{DBLP:conf/osdi/ZhongLCHZL0024,DBLP:conf/sosp/WuLZ0L024,DBLP:conf/osdi/ZhangWLWS0025}:
a TTFT SLO that grows linearly with the prompt length,
$b + a \cdot L$ for a prompt of $L$ tokens,
that is, a fixed budget $b$ plus a per-token budget $a$;
and a fixed TPOT SLO.

We compare the methods at one offered request rate per
global-tier provisioning, the same rate for every method, under a fixed SLO;
\textsection{\ref{sec:eval}} varies the rate and the SLO, with consistent results.
To further measure the methods under different {\kvcache} pressure,
we vary the provisioning of the global tier.
We call the global tier \emph{fully provisioned}
when its capacity is large enough that a request routed to any instance
can usually fetch its session's {\kvcache} from it.
Our testbed's global tier can cache nearly all ongoing sessions,
so a fully provisioned tier always serves the {\kvcache};
in production, however, we do observe under-provisioned global tiers,
with insufficient capacity or bandwidth.

\begin{figure}[!t]
    \begin{minipage}{1\linewidth}
    \centering
    \makebox[\linewidth][c]{%
        \includegraphics[width=1.03\linewidth]{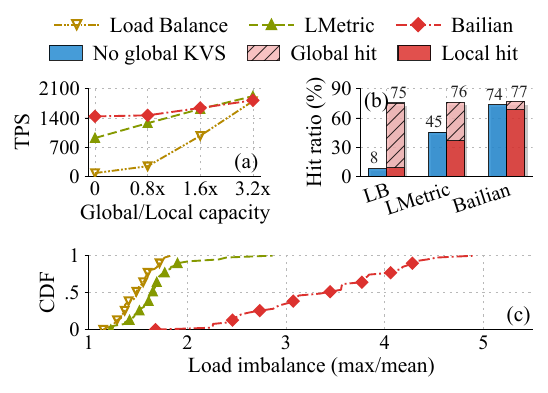}%
    }
    \end{minipage} \\[0pt]
    \begin{minipage}{1\linewidth}
    \caption{\small{%
        (a) TPS within SLO under different global-tier offerings.
        (b) {\kvcache} reuse ratio and its source.
        (c) Distribution of the per-30-second-window load imbalance.
    }}
    \label{fig:motivation-existing-schedulers}
    \end{minipage} \\[0pt]
\end{figure}

\stitle{The {\kvcache}-awareness vs. load balance trade-off. \,}
Existing schedulers still trade {\kvcache} reuse for load balance.
LMetric~\cite{zhang2026simple} reports {\kvcache} reuse ratios comparable to those
of other {\kvcache}-aware schedulers while improving load balance,
on workloads with ideal {\kvcache} reuse ratios typically around 50--70\,\%.
Our agentic serving workload, with an ideal reuse ratio of 82\,\%
(Finding~\ref{finding:reuse}), shows that these two benefits do not always hold together:
without a global tier, LMetric's reuse ratio is only 45\,\%, compared with {\company}'s
74\,\% ({\fig{fig:motivation-existing-schedulers}}\,(b)).
As a result,
without a global tier, LMetric serves only 920 TPS within SLO, 36\,\% below
{\company}'s 1,437 TPS ({\fig{fig:motivation-existing-schedulers}}\,(a)),
even though LMetric achieves better load balance than {\company}'s linear combination
({\fig{fig:motivation-existing-schedulers}}\,(c)).
We measure each instance's load as its average number of running requests
within a 30\,s window.
Load imbalance is the ratio of the maximum load to the mean across the
32 instances~\cite{DBLP:conf/ics/PearceGSSA12,power-of-two}.

Although the ideal reuse ratio of the agentic workload is only about 10 percentage points
above the high end of the non-agentic ones (82\,\% vs.\ 70\,\%),
the difference in prefill computation is far larger:
raising the reuse ratio from 70\,\% to 80\,\% cuts the prompt tokens
that must still be computed by one third.

\stitle{Load balance does not always trade away {\kvcache} reuse. \,}
We do observe opportunities for improvement over policies that favor {\kvcache} reuse:
with a fully provisioned global tier, load balancing need not reduce {\kvcache} reuse.
This is because
the global tier can serve the {\kvcache} missed in the local tier, so even a
pure load-balance method
achieves a high reuse ratio
({\fig{fig:motivation-existing-schedulers}}\,(b)).
As shown in {\fig{fig:motivation-existing-schedulers}}\,(a),
the TPS within SLO of a load-balance-only solution
is within 7\,\% of the best of the three methods.

\stitle{Local hits are still important. \,}
Although {\kvcache} hits from the global tier can recover reuse,
the local reuse is still important:
Load-balance-only and {\company} both reuse about 76\,\% of prefill tokens,
yet their TPS is similar with sufficient global-tier capacity
({\fig{fig:motivation-existing-schedulers}}\,(a, b)),
despite that Load-balance-only has a better load balance:
1.5$\times$ vs.\ 3.3$\times$ imbalance
as shown in {\fig{fig:motivation-existing-schedulers}}\,(c).
This is because
the local hit ratio of Load-balance-only is only 9.1\,\%, vs.\ {\company}'s 68.8\,\%,
requiring more global-tier transfers that consume bandwidth and incur latency.

\takeaway{Scheduling should maintain \uline{load balance} and \uline{{\kvcache} reuse}
while preserving as much \uline{local} reuse as possible.}

\subsection{Session-centric scheduling: challenges and our approach}
\label{sec:solution-v2}

\nospacestitle{Opportunity: session-centric scheduling can reconcile load balance and {\kvcache}-awareness. \,}
Our insight is that routing by session, rather than by request,
can achieve both objectives.
Suppose the scheduler balances sessions across the cluster and routes
each follow-up request to the instance serving its session.
This strategy can balance the load without losing reuse, because the token load
can still be balanced at scale despite the skew of sessions, as characterized in
\textsection{\ref{sec:analyze-session}} (Finding~\ref{finding:balance}).
Note that placing sessions is coarser than placing requests,
which requires further care to balance the load (as we describe in \emph{Challenge\#\,2} below in more detail).
It also keeps (local) {\kvcache} reuse high:
every request except the first reuses the {\kvcache} of earlier
requests in the same session (Finding~\ref{finding:intra}),
and the reuse comes quickly (Finding~\ref{finding:speed}),
so the {\kvcache} usually still resides in the local tier.
The first-turn request is the only exception,
and it loses little performance because its reusable {\kvcache} is
mostly the shared system prompt (Finding~\ref{finding:system-prompt}),
which the global tier serves.
The load on the global tier also stays low, because in the common
case only first-turn requests read from it.

Realizing session-centric routing faces two challenges:

\etitle{\#Challenge\,1: missing session information. \,}
Existing session-centric schedulers like llm-d~\cite{llm-d} require
each request to carry a session identifier (ID),
which is impractical when a provider serves agents, for two reasons.
First, requests carry no session identifier.
The APIs that agents use, e.g., Anthropic's Messages API and OpenAI's
Chat Completions API~\cite{anthropic-api,openai-api}, are stateless:
each request resends the session's full history and carries no
identifier;
stateful variants~\cite{openai-responses} are optional and disabled
under zero-data-retention policies.
Nor can the provider add one:
agents send requests through diverse frameworks
(e.g., Claude Code~\cite{claudecode} and OpenClaw~\cite{openclaw}),
and it is impractical for these frameworks to agree on a session protocol.
Second, the router cannot infer the session from the request's
source: at {\company}, enterprise customers send requests through
global proxies, which anonymize the source (e.g., the IP).

One possible solution is to let the router infer sessions by itself,
e.g., by recording past requests and matching each new request
against the records.
This is impractical:
existing routers keep only per-instance state, i.e., the indicators in
{\fig{fig:general-scheduler}}.
Per-session state, in contrast, scales with the number of active sessions,
which far exceeds the number of instances the router manages,
and must be updated as sessions come and go.
This complicates the router,
and to our knowledge no existing LLM scheduler adopts it.

\begin{figure}[!t]
    \begin{minipage}{1\linewidth}
    \centering
    \includegraphics{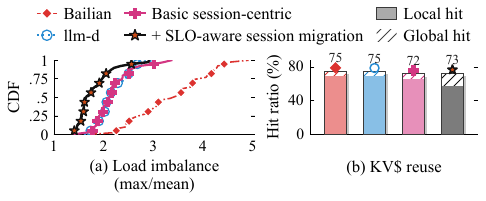}
    \end{minipage} \\[1pt]
    \begin{minipage}{1\linewidth}
    \caption{\small{%
        (a) Per-window load imbalance, as in
        {\fig{fig:motivation-existing-schedulers}}\,(c).
        (b) {\kvcache} reuse ratio and its source.
        Each arm runs at the request rate where its group peaks in TPS within SLO.
    }}
    \label{fig:affinity-load-imbalance}
    \end{minipage} \\[-5pt]
\end{figure}

\etitle{\#Challenge\,2: unconditional sticking still imbalances the load. \,}
Sticking every request to its session's instance can still imbalance
the load, because sessions differ widely in size
(\textsection{\ref{sec:analyze-session}}) and the size is unknown
when the session is first routed.
Suppose several sessions placed on the same instance all turn out to
be long, with many turns:
each follow-up turn returns to the instance with a longer prompt to
prefill, so the instance overloads as the sessions grow.
{\fig{fig:affinity-load-imbalance}} measures this on our testbed.
\mbox{llm-d}~\cite{llm-d} routes requests by session ID,
which we reconstruct offline from the trace and supply with each
request (\textsection{\ref{sec:trace-analyze}}).
Its mean imbalance is still 2.17$\times$,
though it is more balanced than {\company}'s linear combination of
{\kvcache}-awareness and load (3.34$\times$).

\stitle{Basic solution: the turn and the {\kvcache} hit identify the session without per-session state. \,}
To address the first challenge,
we observe that session-centric scheduling leverages session
information to answer two questions per request for routing:
(1) whether the request starts a session and thus should be
load-balanced, and
(2) for a follow-up request, which instance served the session's
previous request, so that the request can stick to it for
{\kvcache}-awareness.
Both can be answered without per-session state.
Specifically,
the first can be answered via the request's turn:
a first-turn request marks the arrival of a new session.
Since
requests carry the session's full history (\textsection{\ref{sec:bg}}),
the router deduces the turn from the number of historical messages
the request carries.
The deduction fails only if the agent drops part of its history,
e.g., by compaction.
The failure is benign:
such a request hits little {\kvcache} anyway, so routing it by load
balance suffices.
The second can be answered via the local {\kvcache} hit:
a follow-up request mostly reuses the {\kvcache} of its own session
(Finding~\ref{finding:intra}), and the reuse comes quickly
(Finding~\ref{finding:speed}),
so the instance with the highest local hit is likely the one that
served the previous request.

\begin{figure}[!t]
\begin{lstlisting}[style=ssched,firstnumber=1]
@\lstregion{(a) stateless session-centric routing}@req = receive()
c = kvc.hit_len(req)   # local KV$ hit, per instance
q = queued_cost()  # sum of prefill_cost in i's queue
prev = instances.argmax(c)  # likely served last turn

if req.turn != 0 and session_cached(req, prev) \
        and (meets_ttft(req, prev)        # (b)
             or none_meets_ttft(req)):
    sched_to = prev                       # stick
else:                                     # balance
    sched_to = instances.rr_argmin(
        key=lambda i: estimate_load(req, i))
req.forward(sched_to)

def session_cached(req, i): # i caches req's session
    return c[i] > HIT_RATIO * est_hit(req)
    # est_hit: the hit expected from req's history
\end{lstlisting}
\vspace{6pt}
\begin{lstlisting}[style=ssched,firstnumber=last,
    backgroundcolor=\color{green!5}]
@\lstregion{(b) SLO checks}@def meets_ttft(req, i):
    # the deadline check
    ttft = estimate_load(req, i) / PREFILL_RATE
    return ttft <= SLACK * ttft_slo(req)

def none_meets_ttft(req):  # the last resort:
    # no instance meets the deadline, so stick
    return not any(meets_ttft(req, i)
                   for i in instances)
\end{lstlisting}
\vspace{6pt}
\begin{lstlisting}[style=ssched,firstnumber=last,
    backgroundcolor=\color{purple!4}]
@\lstregion{(c) load model}@def estimate_load(req, i):
    # i's queued prefill + req's own
    return q[i] + prefill_cost(req, c[i])

def prefill_cost(req, hit): # Eq. @\ref{eq:prefill-cost}@, reusing hit
    L = req.prompt_len; n = L - hit  # new tokens
    return C_LIN * n + C_ATT * n * (L - n / 2)
\end{lstlisting}
\vspace{3pt}
\caption{\small{
    Simplified pseudocode of {\sys}.
    \texttt{C\_LIN} and \texttt{C\_ATT} are constants fitted by profiling the model.
    \texttt{PREFILL\_RATE} (the instance prefill rate) is measured,
    while \texttt{SLACK} and \texttt{HIT\_RATIO} are the two
    hyperparameters.
}}
\label{alg:ssched}
\end{figure}

{\fig{alg:ssched}}\,(a) shows our basic policy.
It assumes PD disaggregation and routes prefill requests only;
the extension to PD colocation is similar (though a little more complex)
and is left to
\textsection{\ref{sec:appendix-coloc}} due to space limitations.
The router first checks whether a request is a first-turn request
(line 5).
If so, it routes the request to the least-loaded instance, which
balances sessions across the cluster (lines 9--11);
otherwise it sends the request to the instance with the
highest {\kvcache} hit (lines 4 and 8)---the instance that likely
executed the session's previous request.
The load in the balance branch is estimated by the model in (c),
which subtracts the request's local hit from its prefill work,
so the routing decision stays {\kvcache}-aware:
a first-turn request still favors the instances that cache its
system prompt.
As shown in {\fig{fig:affinity-load-imbalance}},
the basic policy achieves a load-imbalance curve and {\kvcache} reuse
similar to
session-centric scheduling (llm-d) with session IDs.

Two cases are worth mentioning.
First, a follow-up request whose session has been evicted from the
local tier is treated as a first-turn request, since there is no
local {\kvcache} it can use.
\texttt{session\_cached} detects eviction by comparing the highest
local hit against the hit expected from the request's history
(\texttt{est\_hit}, which excludes the newly appended turn that no
instance has cached);
a large gap indicates that the local tier has evicted the session.
\texttt{HIT\_RATIO} is a pre-configured threshold below one that tolerates small
differences between the estimate and the real hit;
{\sys} is robust to it (\textsection{\ref{sec:eval-abl}}).
Second, {\sys} sticks a follow-up request to its session's instance
only if that instance can finish it within the SLO, which we
elaborate next.

\stitle{Extending the basic policy: SLO-aware session migration via scheduling. \,}
To address the second challenge,
{\sys} sticks a follow-up request to the instance that served its
previous request only if that instance can finish the request within
its SLO.
Otherwise, {\sys} routes the request to the least-loaded instance,
which migrates the session away to prevent SLO violations
due to overloading.
{\fig{fig:affinity-load-imbalance}} shows the effect:
under the same run, migration lowers the basic policy's mean
imbalance from 2.18$\times$ to 1.81$\times$ and leaves no window above 3$\times$.
The migrated request still reuses its {\kvcache}:
the target instance fetches the session's {\kvcache} from the global
tier over RDMA (\ding{194} in {\fig{fig:global-tier}}).
Note that under high load (or bursts), no instance may be able to
meet the deadline.
{\sys} then still routes the request by {\kvcache}-awareness
to add minimal new load to the cluster (line 7).

\texttt{meets\_ttft()} in {\fig{alg:ssched}}\,(b) shows the detailed implementation
of the check.
We use the request's SLO attainment, rather than the instance's load,
as the indicator for migration.
A load-based indicator would also work here, e.g., when the instance's load
exceeds the others' by a margin, we migrate the session off that instance by routing
the request away from it.
However, the load threshold is difficult to set:
too strict a threshold migrates sessions that the instance could still
serve on time and loses their local {\kvcache} reuse,
while too loose a threshold rarely triggers migration.
The SLO sets the threshold directly:
a request migrates exactly when sticking would miss its deadline.
The TTFT deadline (\texttt{ttft\_slo}) is a function set by the
provider, typically linear in the request
length~\cite{DBLP:conf/osdi/ZhongLCHZL0024,DBLP:conf/sosp/WuLZ0L024,DBLP:conf/osdi/ZhangWLWS0025},
and \texttt{SLACK} is a pre-configured hyperparameter that scales
the deadline to leave headroom, to which {\sys} is also robust over a
common range (\textsection{\ref{sec:eval-abl}}).
The check compares the deadline against the estimated prefill time
of the instance's queued requests plus the new request, which we
describe next.

\stitle{Estimating the load. \,}
Both load balancing and the SLO check need a load model that is fast
to compute yet reasonably accurate.
Existing schedulers like Dynamo~\cite{ai-Dynamo} use the number of
tokens to prefill as the load,
which we found inaccurate because the computation of each token
depends on its context~\cite{DBLP:conf/osdi/ZhuGZZZGX0KLWWK25}.
We therefore estimate the load ({\fig{alg:ssched}}\,(c)) in a more
precise form, following prior work~\cite{DBLP:conf/eurosys/ChengLWCC26}:
\begin{equation}
    \label{eq:prefill-cost}
    \mathrm{cost}(n, L) \,=\, c_{\mathrm{linear}}\, n \,+\,
    c_{\mathrm{attn}}\, n\,(L-n/2),
\end{equation}
where $L$ is the request's context (prompt) length and $n$ is the
number of new tokens to prefill.
$c_{\mathrm{linear}}$ and $c_{\mathrm{attn}}$ are coefficients
fitted by profiling,
and the term $n(L-n/2)$ follows from causal (lower-triangular)
attention:
each new token attends to all tokens before it, on average $L-n/2$
of them.
\textsection{\ref{sec:eval-prefilter}} quantifies the contribution of this model.

%% file: body/eval.tex
\section{Performance evaluation}
\label{sec:eval}

\subsection{Evaluation setup}
\label{sec:eval-setup}

\nospacestitle{Testbed. \,}
Our testbed is a cluster of four GPU servers,
each equipped with eight NVIDIA H20 GPUs (96\,GB HBM),
160 CPU cores, and 1280\,GB of host DRAM.
Servers are interconnected with 200\,Gbps RDMA NICs.
Each instance runs the latest
vLLM~\cite{vllm,vllm-code}.
The global {\kvcache} store is based on
Mooncake~\cite{305212} and LMCache~\cite{lmcache},
the state-of-the-art global-tier {\kvcache} store
described in {\fig{fig:global-tier}}.
We tuned each system to its peak performance:
e.g., {\kvcache} fetches from the global tier nearly saturate the RDMA bandwidth.
Unless otherwise mentioned,
we use all instances to evaluate the performance,
and fully provision the global tier;
\textsection{\ref{sec:eval-e2e}} and \textsection{\ref{sec:eval-e2e-pd}}
also evaluate with different global {\kvcache} capacities.

\begin{figure}[!t]
    \vspace{2mm}
    \begin{minipage}{1\linewidth}
    \centering
    \includegraphics{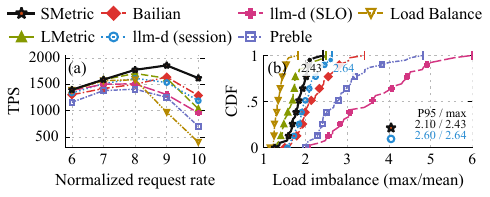}
    \end{minipage} \\[3pt]
    \begin{minipage}{1\linewidth}
    \caption{\small{%
        (a) TPS within SLO and (b) distribution of the per-window
        load imbalance (\textsection{\ref{sec:analysis-methods}})
        on the 32-instance Qwen3-30B-A3B PD-colocated setting.
    }}
    \label{fig:qwen30b-32inst-load-and-tps}
    \end{minipage} \\[-5pt]
\end{figure}

\stitle{Evaluated model. \,}
We evaluate two models, Qwen3-Coder-30B-A3B
and Qwen3-235B-A22B-FP8, both popular models on the market~\cite{DBLP:conf/sosp/Xiang0QYZYZL0025}.
Qwen3-30B can be served with only one GPU, allowing us to
evaluate up to 32 instances.
In contrast, Qwen3-235B requires at least 4 GPUs per instance,
so we can evaluate up to 8 instances.

\stitle{Evaluated workload. \,}
We replay Trace 1 (\textsection{\ref{sec:trace-analyze}}) for all experiments.
Because the requests are collected from a much larger cluster than our testbed,
we sample the requests for the evaluations
following prior work~\cite{zhang2026simple,DBLP:conf/osdi/ZhongLCHZL0024,DBLP:conf/asplos/MiaoSDXL0J24,DBLP:conf/osdi/GujaratiKAHKVM20,traceupscaler},
keeping every request of a session sampled so that its {\kvcache} reuse chain stays complete.
We report the request rate as a multiple of a base rate,
a small fixed fraction of the trace that we cannot disclose due to privacy requirement of {\company}.
One issue with replaying is that the evaluated model's responses
differ from those recorded in the trace.
Such a difference breaks the {\kvcache} reuse pattern in two ways:
the generated response may have a different length,
and a follow-up request's recorded history no longer matches
the tokens the serving instance actually cached.
We fix the issue as follows.
First, we force the evaluated model to generate
exactly as many tokens as the recorded response,
by ignoring the end-of-sequence token and
capping the generation at the recorded length.
Second, we rewrite part of the follow-up requests
with the generated content
to ensure consistent {\kvcache} reuse.

\subsection{End-to-end performance: PD-colocation}
\label{sec:eval-e2e}

\begin{figure}[!t]
    \vspace{2mm}
    \begin{minipage}{1\linewidth}
    \centering
    \includegraphics{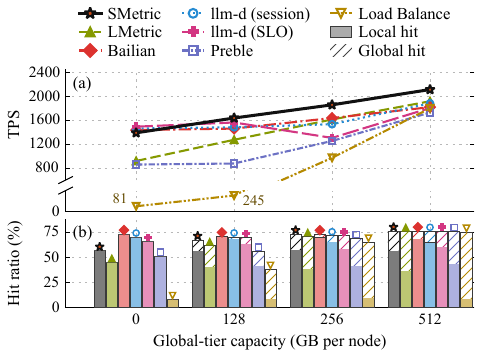}
    \end{minipage} \\[5pt]
    \begin{minipage}{1\linewidth}
    \caption{\small{%
        (a) TPS within SLO and (b) {\kvcache} reuse ratio and its
        source under different global-tier capacities on the 32-instance
        Qwen3-30B-A3B PD-colocated setting.
    }}
    \label{fig:qwen30b-32inst-reuse-by-cpu-cache-allpoints}
    \end{minipage} \\[-10pt]
\end{figure}

\nospacestitle{Compared targets. \,}
We compare {\sys} with the state-of-the-art schedulers described in
\textsection{\ref{sec:analysis-methods}},
as well as other schedulers from well-known open-source projects and
research work.
llm-d (session)~\cite{llm-d} routes requests by session ID;
it is the session-centric baseline described in
\textsection{\ref{sec:solution-v2}}.
We supply the session ID reconstructed offline.
llm-d (SLO) routes by predicted latency:
an online-trained model predicts the TTFT and TPOT the request would
see on each instance,
and the request goes to the instance with the least predicted headroom
among those meeting the SLOs we score against.
Other non-open-source schedulers like the Mooncake scheduler~\cite{305212}
follow a similar rationale.
Preble~\cite{DBLP:conf/iclr/SrivatsaHAL025} filters instances by
{\kvcache} hit before scoring them:
a request whose cached prefix covers more than half of its prompt
goes to the least loaded instance holding that prefix,
and any other request to the instance with the least estimated work,
a linear-combined ({\company}-like) score.
We use its implementation in AIBrix~\cite{aibrix}.

\begin{figure}[!t]
    \begin{minipage}{1\linewidth}
    \centering
    \includegraphics{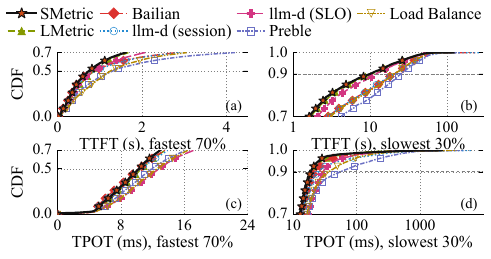}
    \end{minipage} \\[1pt]
    \begin{minipage}{1\linewidth}
    \caption{\small{%
        Per-token latency distributions on the 32-instance
        Qwen3-30B-A3B PD-colocated setting.
        (b, d) use a log scale.
    }}
    \label{fig:qwen30b-32inst-latency-cdf}
    \end{minipage} \\[-15pt]
\end{figure}

\begin{figure}[!t]
    \begin{minipage}{1\linewidth}
    \centering
    \includegraphics{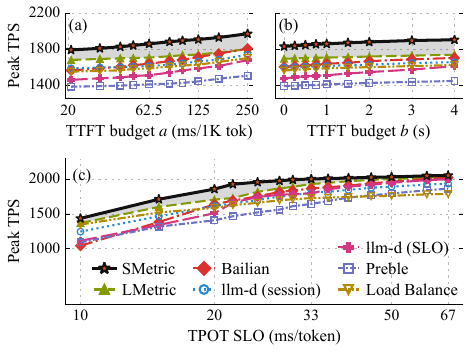}
    \end{minipage} \\[1pt]
    \begin{minipage}{1\linewidth}
    \caption{\small{%
        Sensitivity to SLO configurations on the 32-instance
        Qwen3-30B-A3B PD-colocated setting.
    }}
    \label{fig:qwen30b-32inst-slo-sensitivity}
    \end{minipage} \\[-15pt]
\end{figure}

\begin{figure}[!t]
    \vspace{2mm}
    \begin{minipage}{1\linewidth}
    \centering
    \includegraphics{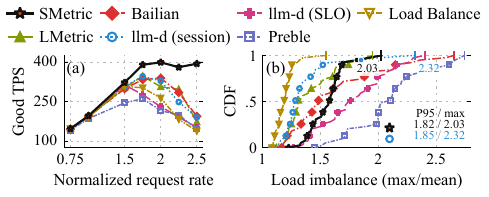}
    \end{minipage} \\[3pt]
    \begin{minipage}{1\linewidth}
    \caption{\small{%
        (a) TPS within SLO and (b) distribution of the per-window
        load imbalance (\textsection{\ref{sec:analysis-methods}})
        on the 8-instance Qwen3-235B-A22B-FP8 PD-colocated setting.
    }}
    \label{fig:qwen235b-8inst-kvs32-load-and-tps}
    \end{minipage} \\[-10pt]
\end{figure}

\stitle{TPS and per-token latency. \,}
{\fig{fig:qwen30b-32inst-load-and-tps}}\,(a) reports the TPS within SLO
(\textsection{\ref{sec:analysis-methods}})
on Qwen3-30B with all 32 instances under the fully provisioned store,
at increasing request rates and a fixed SLO
(TTFT budgets $b = 1$\,s and $a = 62.5$\,ms per 1K prompt tokens;
TPOT within 20\,ms).
{\sys} achieves the highest TPS within SLO:
its peak exceeds the best baseline's by 9\,\%
(1860 vs.\ LMetric's 1708\,tok/s, reached one rate lower).
The improvement over the best baseline grows with load:
within 1\,\% at $6\times$ load, +4\,\% at $8\times$,
+13\,\% at $9\times$, and 26\,\% past saturation ($10\times$),
where every scheduler loses TPS and routing mistakes turn directly
into SLO violations.
The result is consistent across global-tier provisionings
(see {\fig{fig:qwen30b-32inst-reuse-by-cpu-cache-allpoints}}\,(a)),
except for setups without any global {\kvcache} store.
This is acceptable, because deploying the global store for serving agents
is now common practice.
The per-token latency in {\fig{fig:qwen30b-32inst-latency-cdf}}
shows similar results: {\sys} has the lowest TTFT at every reported
percentile and the lowest P90 TPOT.

{\sys} serves better because it keeps local {\kvcache} hits and
load balance simultaneously, without trading away the overall reuse ratio.
{\fig{fig:qwen30b-32inst-reuse-by-cpu-cache-allpoints}}\,(b) shows that
{\sys} matches the highest reuse ratio ({\company}'s, 73\,\% for both)
and serves most of it from the local tier (58\,\%).
{\fig{fig:qwen30b-32inst-load-and-tps}}\,(b) shows that it does so
without trading load balance.

\stitle{Sensitivity to the SLO. \,}
{\fig{fig:qwen30b-32inst-slo-sensitivity}} further compares the
TPS within SLO under various SLO setups, as different applications may have different
SLO requirements.
{\sys} is robust across all SLO setups and achieves
the best TPS among all baselines:
it is
7--10\,\% above the best baseline across different TTFT budgets,
and 5--12\,\% at TPOT SLOs of 10--22\,ms.
As the TPOT SLO becomes looser, the gap between {\sys} and others
narrows, because a loose SLO can tolerate more computation
or queuing, but {\sys} can still achieve a lower latency
thanks to better scheduling.

\begin{figure}[!t]
    \vspace{2mm}
    \begin{minipage}{1\linewidth}
    \centering
    \includegraphics{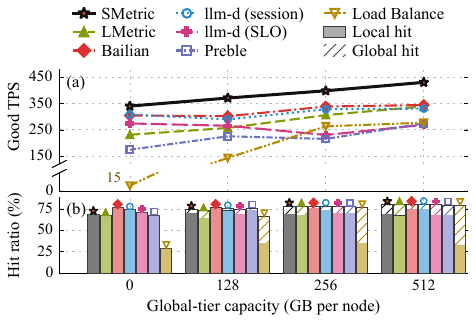}
    \end{minipage} \\[5pt]
    \begin{minipage}{1\linewidth}
    \caption{\small{%
        (a) TPS within SLO and (b) {\kvcache} reuse ratio and its
        source under different global-tier capacities on the 8-instance
        Qwen3-235B-A22B-FP8 PD-colocated setting.
    }}
    \label{fig:qwen235b-8inst-reuse-by-cpu-cache-allpoints}
    \end{minipage} \\[-5pt]
\end{figure}

\begin{figure}[!t]
    \begin{minipage}{1\linewidth}
    \centering
    \includegraphics{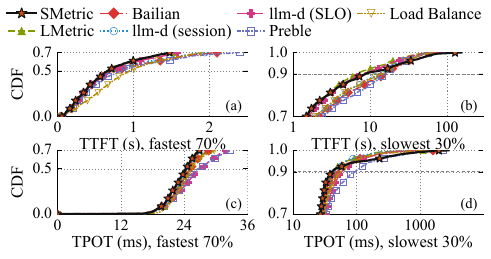}
    \end{minipage} \\[1pt]
    \begin{minipage}{1\linewidth}
    \caption{\small{%
        Per-token latency distributions on the 8-instance
        Qwen3-235B-A22B-FP8 PD-colocated setting.
        (b, d) use a log scale.
    }}
    \label{fig:qwen235b-8inst-kvs32-latency-cdf}
    \end{minipage} \\[-5pt]
\end{figure}

\stitle{Performance on a larger model. \, }
To see how {\sys} performs under larger models where
each instance requires multiple GPUs,
we repeat the evaluation of the 30B model with Qwen3-235B on eight instances,
with the TPOT SLO relaxed to 33\,ms, as the larger model decodes slower.
As shown in \fig{fig:qwen235b-8inst-kvs32-load-and-tps}\,(a),
the results are consistent with those on Qwen3-30B:
{\sys} again achieves the highest TPS within SLO,
with a peak 15\,\% above the best baseline's
(399 vs.\ LMetric's 346\,tok/s),
a gap that widens to 28\,\% as the load grows ($2.25\times$),
the lowest median TTFT and P90 TPOT
({\fig{fig:qwen235b-8inst-kvs32-latency-cdf}}),
and consistent results across different SLO configurations
({\fig{fig:qwen235b-8inst-kvs32-slo-sensitivity}} in the appendix).
Serving the large model is similar to serving smaller models
from the perspective of scheduling, because
the difference is at the instance level,
whereas scheduling routes requests across instances.
{\sys} still keeps the {\kvcache} reuse and the load balance as before:
it matches the highest reuse ratio (78\,\%, the same as LMetric's)
and serves 70\,\% of the prompt tokens from the local tier
({\fig{fig:qwen235b-8inst-reuse-by-cpu-cache-allpoints}}\,(b)),
and its load imbalance stays close to LMetric's
({\fig{fig:qwen235b-8inst-kvs32-load-and-tps}}\,(b));
only Load-balance-only, which optimizes that quantity directly, is
more balanced, and it serves 34\,\% less TPS at this rate
(23\,\% peak to peak) because it reuses
{\kvcache} mostly from the global tier
({\fig{fig:qwen235b-8inst-reuse-by-cpu-cache-allpoints}}\,(b)).

Nevertheless, we do make two observations on large models.
First, the largest performance gap widens (15\,\% vs.\ 9\,\% on Qwen3-30B) because the
larger model puts more pressure on the global tier, which makes it
more important for the scheduler's decision to favor local reuse.
Second, llm-d (session) appears more balanced than {\sys} in
{\fig{fig:qwen235b-8inst-kvs32-load-and-tps}}\,(b),
as its curve stays left of ours up to the P90 window.
Despite this, {\sys} can still achieve a better TPS because,
among the {\kvcache}-aware schedulers, it has a lower tail imbalance:
the P99 max/mean imbalance of llm-d (session) reaches 2.20$\times$ and its worst 2.32$\times$,
against 1.99$\times$ (P99) and 2.03$\times$ for {\sys}, respectively.
This tail imbalance is critical to the SLO attainment,
because the requests queued at an overloaded instance
are the ones that miss the SLO.
{\sys} migrates the session away from such an instance
(\textsection{\ref{sec:solution-v2}}),
which reduces the latency increase under extreme imbalance.

\begin{figure}[!t]
    \vspace{2mm}
    \begin{minipage}{1\linewidth}
    \centering
    \includegraphics{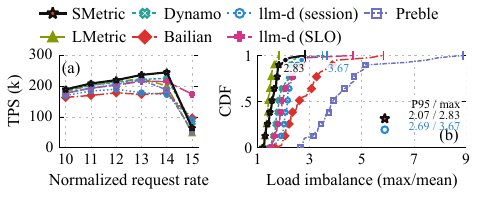}
    \end{minipage} \\[3pt]
    \begin{minipage}{1\linewidth}
    \caption{\small{%
        (a) Prefill TPS within SLO and (b) distribution of the per-window
        load imbalance (\textsection{\ref{sec:analysis-methods}})
        on the 32-instance Qwen3-30B-A3B PD-disaggregated setting.
    }}
    \label{fig:qwen30b-32inst-prefill-load-and-tps}
    \end{minipage} \\[-0pt]
\end{figure}

\subsection{End-to-end performance: PD-disaggregation}
\label{sec:eval-e2e-pd}

\nospacestitle{Evaluation setup and compared targets. \,}
We next evaluate {\sys} under PD disaggregation.
Since {\kvcache} reuse only affects prefill,
we deploy 32 prefill instances of Qwen3-30B,
run only the prefill phase, and report the
\emph{prefill TPS} within SLO---the number of prompt tokens whose
{\kvcache} is produced per second, computed or reused from a
hit---and the latency results (TTFT).
We use only the single-card model (30\,B) to evaluate at a larger scale;
the results on the multi-card model (235\,B) are similar.

The compared targets are the same as under PD colocation,
except that we further evaluate Dynamo~\cite{ai-Dynamo},
a state-of-the-art system, which optimizes upon Load-balance-only:
it routes a request to the instance with the fewest tokens left to
prefill, including the queued tokens of other requests
and the new tokens of the request to route considering {\kvcache} hits.

\begin{figure}[!t]
    \vspace{2mm}
    \begin{minipage}{1\linewidth}
    \centering
    \includegraphics{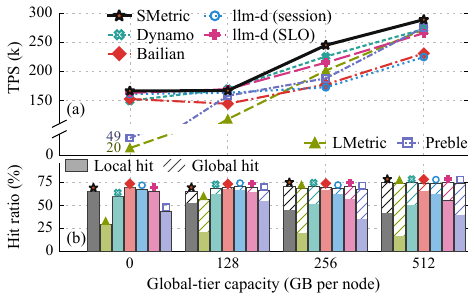}
    \end{minipage} \\[3pt]
    \begin{minipage}{1\linewidth}
    \caption{\small{%
        (a) Prefill TPS within SLO and (b) {\kvcache} reuse ratio and its
        source under different global-tier capacities on the 32-instance
        Qwen3-30B-A3B PD-disaggregated setting.
    }}
    \label{fig:qwen30b-32inst-prefill-reuse-by-cpu-cache-allpoints}
    \end{minipage} \\[-5pt]
\end{figure}

\begin{figure}[!t]
    \begin{minipage}{1\linewidth}
    \centering
    \includegraphics{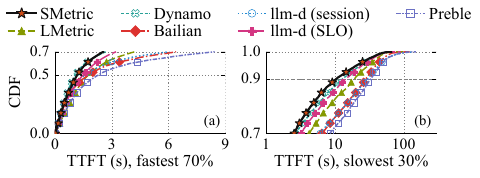}
    \end{minipage} \\[1pt]
    \begin{minipage}{1\linewidth}
    \caption{\small{%
        TTFT distributions on the 32-instance
        Qwen3-30B-A3B PD-disaggregated setting
        for (a) common and (b) tail requests.
        (b) uses a log scale.
    }}
    \label{fig:qwen30b-32inst-prefill-ttft-cdf}
    \end{minipage} \\[-5pt]
\end{figure}

\stitle{Prefill TPS and TTFT. \,}
{\fig{fig:qwen30b-32inst-prefill-reuse-by-cpu-cache-allpoints}}\,(a)
compares the prefill TPS across global-tier provisionings, and
{\fig{fig:qwen30b-32inst-prefill-load-and-tps}}\,(a)
evaluates how the TPS changes with the request rate under the
fully provisioned store.
{\sys} still achieves the highest prefill TPS at three of the four
provisionings, 2--9\,\% above the best-performing baseline;
only with the 128\,GB store does it slightly underperform Dynamo and
llm-d (SLO), by 2\,\%.
The peak TPS achieved by {\sys} across request rates exceeds
Dynamo's peak by 9\,\%
(245 vs.\ 226\,ktok/s, {\fig{fig:qwen30b-32inst-prefill-load-and-tps}}\,(a)).
The latency results (TTFT) in {\fig{fig:qwen30b-32inst-prefill-ttft-cdf}} show similar
results:
{\sys} delivers the lowest tail TTFT, with a P90 19\,\% lower than
Dynamo's, the closest baseline, and with a close median.

We found that balanced schedulers matter more for TPS under
PD disaggregation: Dynamo and LMetric now outperform or are similar to the
{\kvcache}-aware scheduler (\company).
The reason is that prefill is more compute-intensive,
so an imbalanced load directly affects the prefill TPS,
whereas decode can run the queued requests in a larger batch
to amortize the imbalance cost under PD colocation.
Nevertheless, achieving similar {\kvcache} reuse with load balancing,
as {\sys} does, is still beneficial:
with the 256\,GB store it matches the best baselines' reuse ratio
(71\,\%,
{\fig{fig:qwen30b-32inst-prefill-reuse-by-cpu-cache-allpoints}}\,(b))
with a load nearly as balanced as LMetric's:
averaged over the 30\,s windows, the busiest prefill instance holds
1.6$\times$ as many running and waiting requests as the average instance,
against 1.9$\times$ under Dynamo and 2.8$\times$ under
{\company} ({\fig{fig:qwen30b-32inst-prefill-load-and-tps}}\,(b)).

\begin{figure}[!t]
    \begin{minipage}{1\linewidth}
    \centering
    \includegraphics{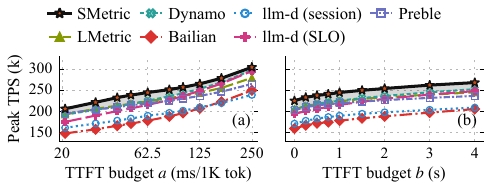}
    \end{minipage} \\[1pt]
    \begin{minipage}{1\linewidth}
    \caption{\small{%
        Sensitivity to SLO configurations on the 32-instance
        Qwen3-30B-A3B PD-disaggregated setting.
    }}
    \label{fig:qwen30b-32inst-prefill-slo-sensitivity}
    \end{minipage} \\[-15pt]
\end{figure}

\stitle{Sensitivity to the SLO. \,}
{\fig{fig:qwen30b-32inst-prefill-slo-sensitivity}} reports the prefill
TPS within SLO under various SLO setups.
The results are similar to those under PD colocation:
{\sys} still achieves the best TPS:
6--10\,\% above the best baseline across different fixed TTFT setups
and 2--9\,\% across different per-token budgets.

We also repeat this evaluation on an open agentic-coding trace,
the Codex agent traces on SWE-Bench Pro~\cite{codex-swebenchpro-trace},
and reach the same conclusions:
{\sys} achieves the highest peak prefill TPS, 10\,\% above the best
baseline, and the lowest TTFT
(\textsection{\ref{sec:appendix-codex}}).

\subsection{Ablation study}
\label{sec:eval-prefilter}

\begin{figure}[!t]
    \begin{minipage}{1\linewidth}
    \centering
    \includegraphics[width=\linewidth]{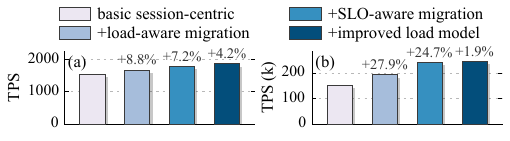}
    \end{minipage} \\[3pt]
    \begin{minipage}{1\linewidth}
    \caption{\small{%
        Ablation study of {\sys} on (a) the PD-colocated and (b) the
        prefill instances of the
        PD-disaggregated 32-instance Qwen3-30B-A3B setups.
    }}
    \label{fig:qwen30b-32inst-mechanism-ablation}
    \end{minipage} \\[-20pt]
\end{figure}

\noindent
This section conducts an ablation study on the routing designs of {\sys}.
We use the 32-instance Qwen3-30B-A3B setups above, under PD
colocation and PD disaggregation,
each at the request rate where {\sys} achieves peak TPS.

As shown in {\fig{fig:qwen30b-32inst-mechanism-ablation}},
our starting point is \emph{basic session-centric},
which only uses the request's turn and its local {\kvcache} hit
(lines 2, 4 and 5 of {\fig{alg:ssched}}) to distinguish sessions
and realize basic session-centric scheduling
(similar to llm-d (session)).
First, adding overloading migration, i.e., replacing the SLO-aware
migration on lines 6--7 with a load-aware filter---a follow-up
request leaves its session's instance when that instance's load
(queued prefill plus ongoing decode) exceeds the cluster mean by a
margin---improves the TPS by 8.8\,\% and 27.9\,\% under colocation
and disaggregation, respectively,
confirming that session-centric scheduling alone cannot handle
overloading cases.
Second, replacing the load-aware filter with our SLO-aware migration
(lines 16--23 in {\fig{alg:ssched}})
further improves the TPS by 7.2\,\% and 24.7\,\%,
showing that the SLO is a more accurate migration signal for TPS.
Finally, replacing the load estimate in the SLO calculation from
token counts to our cost model described in
\textsection{\ref{sec:solution-v2}} (Eq.~\ref{eq:prefill-cost})
increases the TPS by 4.2\,\% and 1.9\,\%,
showing that a more accurate performance model improves scheduling.
Note that using a performance model alone (e.g., llm-d (SLO)) is
still insufficient, as we have evaluated in
\textsection{\ref{sec:eval-e2e}} and \textsection{\ref{sec:eval-e2e-pd}}.

\subsection{Sensitivity to the two routing hyperparameters}
\label{sec:eval-abl}

\begin{figure}[!t]
    \vspace{2mm}
    \centering
    \includegraphics[width=\linewidth]{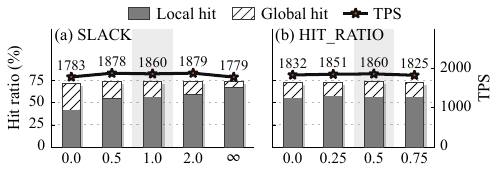}
    \\[3pt]
    \begin{minipage}{1\linewidth}
    \caption{\small{%
        Sensitivity to the two hyperparameters of {\sys} on the 32-instance
        Qwen3-30B-A3B PD-colocated setup.
    }}
    \label{fig:qwen30b-32inst-parameter-ablation}
    \end{minipage} \\[-10pt]
\end{figure}

\noindent
{\sys} uses two hyperparameters in {\fig{alg:ssched}}:
\texttt{SLACK} controls to what extent it can tolerate SLO violations and
\texttt{HIT\_RATIO} determines when the local cache is treated as evicted for the request.
{\fig{fig:qwen30b-32inst-parameter-ablation}} conducts a sensitivity analysis
on these parameters.
{\sys} is robust to both as long as they are set to reasonable values:
TPS within SLO moves by 1.0\,\% over \texttt{SLACK} from 0.5 to 2 and
by 1.9\,\% over the whole range of \texttt{HIT\_RATIO},
except for the improper values, e.g.,
setting \texttt{SLACK} to $0$ disables session stickiness (every
follow-up request migrates), while
setting it to $\infty$ completely ignores the SLO checks.

%% file: body/related.tex
\section{Related Work}
\label{sec:related}

\nospacestitle{LLM request scheduling. \,}
Recent work schedules LLM requests with {\kvcache}-aware, SLO-aware, load-balancing, fairness, and output-length-aware policies~\cite{DBLP:conf/iclr/SrivatsaHAL025,305212,DBLP:conf/osdi/SunHZXZL024,DBLP:journals/corr/abs-2507-17769,DBLP:journals/corr/abs-2504-08784,DBLP:journals/corr/abs-2504-09285,skywalker,ascendra,vtc,dontstopmenow}.
{\sys} builds on these policies and derives agent-session hints for the scheduler without
requiring LLM requests to provide session information.
Unlike systems that co-design serving with agent applications and require applications to expose
inter-request structure~\cite{DBLP:conf/osdi/LinHZ00CQ24,DBLP:journals/corr/abs-2502-13965,pythia,aragog},
{\sys} operates entirely on the provider side and treats agents as unmodified clients,
matching a typical serving scenario today.
AgentKV~\cite{agentkv} also targets agentic serving but learns agent execution to
decide which {\kvcache} to keep; {\sys} needs no learning and infers the session from two
indicators the router already holds.

\stitle{{\kvcache} reuse and optimizations. \,}
Reusing {\kvcache} across requests is well studied,
and both industry and academia have built mature reuse mechanisms~\cite{DBLP:conf/usenix/GaoHSKJDYYZ24,DBLP:conf/mlsys/GimCLSK024,vllm,CacheBlend,DBLP:conf/acl/YeTHL24,sglang,DBLP:journals/corr/abs-2312-05516,unicache},
We build on these mechanisms
and focus on scheduling to better utilize them.
Other work further reduces the memory footprint and transfer cost of {\kvcache} with system--algorithm co-designs:
CacheGen compresses and streams KV states~\cite{DBLP:conf/sigcomm/LiuLCRHZDY0AMHH24,DBLP:journals/corr/abs-2507-11507,DBLP:journals/corr/abs-2602-00328,adaptcache,shadowserve},
and quantization and compression methods reduce {\kvcache} memory pressure~\cite{DBLP:conf/icml/LiuYJZXBC024,DBLP:conf/nips/HooperKMMSKG24,DBLP:conf/nips/LiHYVLYCLC24,DBLP:journals/corr/abs-2406-02069,DBLP:journals/corr/abs-2412-03131}.
These techniques are orthogonal to {\sys}.

\stitle{Optimizing LLM serving. \,}
Scheduling is one part of the serving stack,
and much work targets its other layers,
including efficient attention algorithms~\cite{deepseek-v3,deepseek-v4},
optimized attention kernels and KV layouts~\cite{DBLP:conf/mlsys/00010LLZW0KGKC25,DBLP:conf/asplos/PrabhuNMRP25,DBLP:conf/asplos/KamathPM0RP25},
elastic resource scaling~\cite{DBLP:conf/sosp/Xiang0QYZYZL0025,DBLP:conf/osdi/ZhangWLWS0025},
and many others~\cite{DBLP:conf/mlsys/KangBHKRR25,DBLP:journals/corr/abs-2501-13111,vllm,taichi,helix,adaserve,aqua}.
They are orthogonal to {\sys}, which is at an upper layer,
and we believe that as long as LLMs still reuse {\kvcache},
our techniques and findings are effective.

%% file: body/concl.tex
\section{Conclusion}
\label{sec:concl}

\noindent
Agentic serving demands high cluster-wide TPS
and makes {\kvcache} reuse dominant,
requiring schedulers to be both load-balanced and {\kvcache}-aware.
{\sys} achieves the best of both worlds by first routing
requests by session,
with session information deduced statelessly at the router,
and further deriving an SLO-aware migration scheme to prevent
imbalance caused by unpredictable session growth.
It improves the peak cluster TPS by
9--15\,\% under prefill-decode colocation and the peak prefill TPS by
9\,\% under disaggregation over state-of-the-art schedulers,
at lower latency.

%% file: body/appendix.tex
\clearpage
\appendix

\section{Session reconstruction from anonymized trace}
\label{sec:appendix-session}

\noindent
As the
requests collected in our traces have no session identifier,
we reconstruct sessions offline to simplify analysis and
to compare with session ID-based scheduling methods.

Specifically, each request in our trace contains its
arrival time,
an opaque user ID (a user may be one person or a whole organization),
the prompt as a sequence of 512-token block hashes (the trace carries no prompt text),
the input and output lengths,
the turn number within its session,
and what triggers the request,
e.g., whether the request is a compaction request or a normal one sent by the user or the agent.

To decide which session a request belongs to,
the observation is that one request is the follow up of another from the same user
if:
(1) its turn is larger than the other by 1 and
(2) its hashed block sequence starts with that of the other,
because each turn extends the prompt of its parent with the model's reply and
the tool output or the next user prompt.
As a result, we can simply link each request to the earlier request of the same user
that satisfies both conditions,
chain requests together, and finally assign requests in the same chain the same session ID.

\section{{\sys} under PD colocation}
\label{sec:appendix-coloc}

\noindent
{\fig{alg:ssched-coloc}} presents the pseudocode of {\sys} under PD colocation.
Its high-level structure is similar to the PD-disaggregated version in {\fig{alg:ssched}},
with two main PD-colocation-aware changes.

First, we further select the instances that meet the TPOT SLO (lines 5 and 9)
before deciding where to route,
because a colocated instance also executes decodes,
which the prefill of the new request delays.
If no instance meets the TPOT SLO, the check is skipped (line 6).
Note that the check only considers the decodes in flight at the instance,
because the number of tokens the new request will decode is unknown.
Second, we adopt the multiplicative score of LMetric~\cite{zhang2026simple}
to select the target of load balancing and session migration (lines 12--13),
because in our setup it is the most effective score that considers
prefill and decode in a unified way.

\begin{figure}[!t]
\begin{lstlisting}[style=ssched,firstnumber=1,numbersep=9pt]
@\lstregion{(a) stateless session-centric routing}@req = receive()
c = kvc.hit_len(req)   # local KV$ hit, per instance
q = queued_cost()  # sum of prefill_cost in i's queue
prev = instances.argmax(c)  # likely served last turn
@\lstchg@cand = [i for i in instances if meets_tpot(req, i)]
@\lstchg@if not cand: cand = instances  # (b) none passes

if req.turn != 0 and session_cached(req, prev) \
@\lstchg@        and meets_ttft(req, prev) \
@\lstchg@        and prev in cand:                 # (b)
    sched_to = prev                       # stick
else:                                     # balance
@\lstchg@    sched_to = cand.rr_argmin(key=lambda i:
@\lstchg@        estimate_load(req, i) * i.num_requests)
req.forward(sched_to)

def session_cached(req, i): # i caches req's session
    return c[i] > HIT_RATIO * est_hit(req)
    # est_hit: the hit expected from req's history
\end{lstlisting}
\vspace{6pt}
\begin{lstlisting}[style=ssched,firstnumber=last,numbersep=9pt,
    backgroundcolor=\color{green!5}]
@\lstregion{(b) SLO checks}@def meets_ttft(req, i):
    # the deadline check
    ttft = estimate_load(req, i) / PREFILL_RATE
    return ttft <= SLACK * ttft_slo(req)

@\lstchg@def meets_tpot(req, i):
@\lstchg@    # i's live decodes stay within their TPOT SLO
@\lstchg@    stall = prefill_cost(req, c[i]) / PREFILL_RATE
@\lstchg@    return all(stall <= slo_slack(d)
@\lstchg@               for d in i.live_decodes())
\end{lstlisting}
\vspace{6pt}
\begin{lstlisting}[style=ssched,firstnumber=last,numbersep=9pt,
    backgroundcolor=\color{purple!4}]
@\lstregion{(c) load model}@def estimate_load(req, i):
    # i's queued prefill + req's own
    return q[i] + prefill_cost(req, c[i])

def prefill_cost(req, hit): # Eq. @\ref{eq:prefill-cost}@, reusing hit
    L = req.prompt_len; n = L - hit  # new tokens
    return C_LIN * n + C_ATT * n * (L - n / 2)
\end{lstlisting}
\vspace{3pt}
\caption{\small{
    Pseudocode of {\sys} under PD colocation.
    The bars (\lstbar) mark the lines that differ from {\fig{alg:ssched}}.
    The constants and hyperparameters are the same as {\fig{alg:ssched}}.
}}
\label{alg:ssched-coloc}
\end{figure}

\section{SLO sensitivity on the larger model}
\label{sec:appendix-235b}

\begin{figure}[!b]
    \begin{minipage}{1\linewidth}
    \centering
    \includegraphics{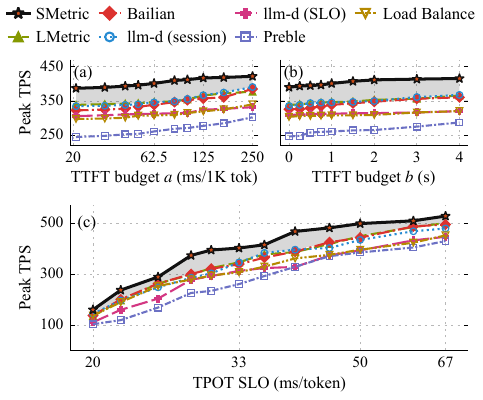}
    \end{minipage} \\[1pt]
    \begin{minipage}{1\linewidth}
    \caption{\small{%
        Sensitivity to SLO configurations on the 8-instance
        Qwen3-235B-A22B-FP8 PD-colocated setting.
    }}
    \label{fig:qwen235b-8inst-kvs32-slo-sensitivity}
    \end{minipage} \\[-5pt]
\end{figure}

\noindent
{\fig{fig:qwen235b-8inst-kvs32-slo-sensitivity}} reports the
sensitivity to the SLO on Qwen3-235B
(\textsection{\ref{sec:eval-e2e}}).
{\sys} achieves the best performance at various SLO setups:
by 8--16\,\% across the fixed and per-token TTFT budgets
and by 4.5--24\,\% across TPOT SLOs of 20--67\,ms.

\clearpage
\section{Results on an open agentic-coding trace}
\label{sec:appendix-codex}

\noindent
To confirm that the PD-disaggregation results
(\textsection{\ref{sec:eval-e2e-pd}}) are not specific to our trace,
we repeat them on an open trace,
the Codex agent traces on SWE-Bench Pro~\cite{codex-swebenchpro-trace},
which record 610 sessions (20,230 requests) of the Codex coding agent.
The trace carries every prompt and response length but no timestamps,
so we sample the think time between turns from the inter-call delays
reported with the dataset.
Its sessions are longer than ours (33 turns on average vs.\ 2.3)
and reuse more of their prompts
(94\,\% of the prompt tokens are cache hits, vs.\ about 80\,\% in ours).
The setup is otherwise that of \textsection{\ref{sec:eval-e2e-pd}}.

The results are consistent with \textsection{\ref{sec:eval-e2e-pd}}.
{\sys} achieves the highest peak prefill TPS, 10\,\% above Dynamo, the
best baseline
({\fig{fig:qwen30b-32inst-prefill-codex-load-and-tps}}\,(a)),
and the most balanced load after LMetric, which gives up {\kvcache}
reuse for balance
({\fig{fig:qwen30b-32inst-prefill-codex-load-and-tps}}\,(b)):
the busiest instance holds 2.5$\times$ as many requests as the average
instance, against 3.1$\times$ or more under the other baselines.
It matches or exceeds the best baseline at every global-tier capacity
({\fig{fig:qwen30b-32inst-prefill-codex-reuse-by-cpu-cache-allpoints}})
and across TTFT budgets
({\fig{fig:qwen30b-32inst-prefill-codex-slo-sensitivity}}),
and delivers the lowest TTFT at both the median and the tail
({\fig{fig:qwen30b-32inst-prefill-codex-ttft-cdf}}).
LMetric and {\company} collapse at a lower request rate than on our
trace, for different reasons: LMetric serves most follow-ups from the
global tier
({\fig{fig:qwen30b-32inst-prefill-codex-reuse-by-cpu-cache-allpoints}}\,(b)),
and the longer prompts make each such transfer costlier;
{\company} weighs {\kvcache} hits far above load, so it keeps sending a
session's follow-ups to its home even when the home is busy, and the
long sessions keep it busy.

\noindent
\begin{minipage}{\linewidth}
    \centering
    \includegraphics{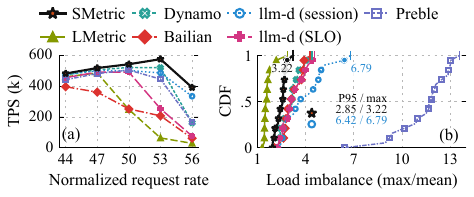}\\[3pt]
    \captionof{figure}{\small{%
        (a) Prefill TPS within SLO and (b) distribution of the per-window
        load imbalance, as in
        {\fig{fig:qwen30b-32inst-prefill-load-and-tps}},
        on the 32-instance Qwen3-30B-A3B PD-disaggregated setting,
        replaying the Codex trace.
    }}
    \label{fig:qwen30b-32inst-prefill-codex-load-and-tps}
\end{minipage}
\vspace{10pt}

\noindent
\begin{minipage}{\linewidth}
    \centering
    \includegraphics{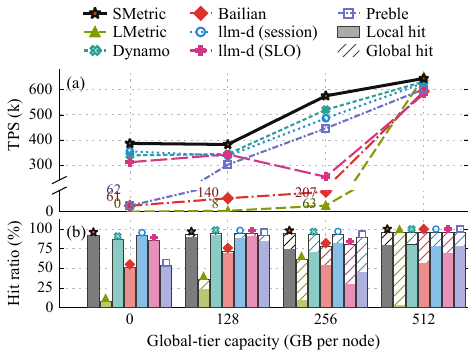}\\[3pt]
    \captionof{figure}{\small{%
        (a) Prefill TPS within SLO and (b) {\kvcache} reuse ratio and its
        source under different global-tier capacities on the 32-instance
        Qwen3-30B-A3B PD-disaggregated setting, replaying the Codex trace.
    }}
    \label{fig:qwen30b-32inst-prefill-codex-reuse-by-cpu-cache-allpoints}
\end{minipage}
\vspace{10pt}

\noindent
\begin{minipage}{\linewidth}
    \centering
    \includegraphics{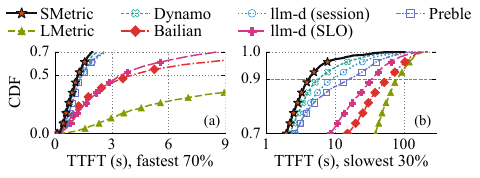}\\[3pt]
    \captionof{figure}{\small{%
        TTFT distributions on the 32-instance
        Qwen3-30B-A3B PD-disaggregated setting
        for (a) common and (b) tail requests, replaying the Codex trace.
        (b) uses a log scale.
    }}
    \label{fig:qwen30b-32inst-prefill-codex-ttft-cdf}
\end{minipage}
\vspace{10pt}

\noindent
\begin{minipage}{\linewidth}
    \centering
    \includegraphics{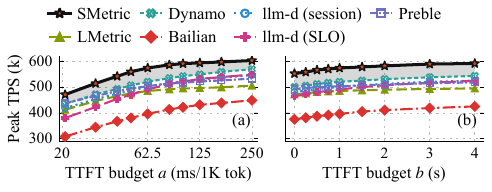}\\[3pt]
    \captionof{figure}{\small{%
        Sensitivity to SLO configurations on the 32-instance
        Qwen3-30B-A3B PD-disaggregated setting, replaying the Codex trace.
    }}
    \label{fig:qwen30b-32inst-prefill-codex-slo-sensitivity}
\end{minipage}
\vspace{10pt}

%% file: paper.bbl
\begin{thebibliography}{10}

\bibitem{298679}
Amey Agrawal, Nitin Kedia, Ashish Panwar, Jayashree Mohan, Nipun Kwatra,
  Bhargav Gulavani, Alexey Tumanov, and Ramachandran Ramjee.
\newblock Taming {Throughput-Latency} tradeoff in {LLM} inference with
  {Sarathi-Serve}.
\newblock In {\em 18th USENIX Symposium on Operating Systems Design and
  Implementation (OSDI 24)}, pages 117--134, Santa Clara, CA, July 2024. USENIX
  Association.

\bibitem{aibrix}
Aibrix.
\newblock \burl{https://github.com/vllm-project/aibrix}, 2025.

\bibitem{aigw}
Aigw.
\newblock \burl{https://github.com/aigw-project/aigw}, 2025.

\bibitem{claudecodeagents2024}
{Anthropic}.
\newblock Claude code: Sub-agents and agent teams.
\newblock \url{https://docs.anthropic.com/en/docs/claude-code/sub-agents},
  2024.

\bibitem{anthropic2025ccompiler}
{Anthropic}.
\newblock Building a {C} compiler with {Claude}.
\newblock \url{https://www.anthropic.com/engineering/building-c-compiler},
  2025.

\bibitem{claudecode}
{Anthropic}.
\newblock Claude code.
\newblock \url{https://code.claude.com/docs/en/overview}, 2026.

\bibitem{anthropic-api}
{Anthropic}.
\newblock Claude platform docs: Using the {Messages} {API}.
\newblock
  \url{https://platform.claude.com/docs/en/build-with-claude/working-with-messages},
  2026.

\bibitem{claude-opus4-outage}
{Anthropic}.
\newblock Elevated error rates on {Opus 4.6}.
\newblock Anthropic Status Page, 2026.

\bibitem{bai2026aiagentsspendmoney}
Longju Bai, Zhemin Huang, Xingyao Wang, Jiao Sun, Rada Mihalcea, Erik
  Brynjolfsson, Alex Pentland, and Jiaxin Pei.
\newblock How do ai agents spend your money? analyzing and predicting token
  consumption in agentic coding tasks, 2026.

\bibitem{DBLP:journals/corr/abs-2406-02069}
Zefan Cai, Yichi Zhang, Bofei Gao, Yuliang Liu, Tianyu Liu, Keming Lu, Wayne
  Xiong, Yue Dong, Baobao Chang, Junjie Hu, and Wen Xiao.
\newblock Pyramidkv: Dynamic {KV} cache compression based on pyramidal
  information funneling.
\newblock {\em CoRR}, abs/2406.02069, 2024.

\bibitem{DBLP:journals/corr/abs-2504-08784}
Siyuan Chen, Zhipeng Jia, Samira Khan, Arvind Krishnamurthy, and Phillip~B.
  Gibbons.
\newblock Slos-serve: Optimized serving of multi-slo llms.
\newblock {\em CoRR}, abs/2504.08784, 2025.

\bibitem{DBLP:conf/eurosys/ChengLWCC26}
Rongxin Cheng, Yuxin Lai, Xingda Wei, Rong Chen, and Haibo Chen.
\newblock {KUNSERVE:} parameter-centric memory management for efficient memory
  overloading handling in {LLM} serving.
\newblock In Antonio Barbalace, Luo Mai, Roxana Geambasu, and Peter~R.
  Pietzuch, editors, {\em Proceedings of the 21st European Conference on
  Computer Systems, EuroSys 2026, McEwan Hall/The University of Edinburgh,
  Edinburgh, Scotland, UK, April 27-30, 2026}, pages 1244--1260. {ACM}, 2026.

\bibitem{aragog}
Yinwei Dai, Zhuofu Chen, Anand Iyer, and Ravi Netravali.
\newblock Aragog: Just-in-time model routing for scalable serving of agentic
  workflows.
\newblock {\em CoRR}, abs/2511.20975, 2025.

\bibitem{deepseek-v3}
DeepSeek{-}AI.
\newblock Deepseek-v3 technical report.
\newblock {\em CoRR}, abs/2412.19437, 2024.

\bibitem{deepseek-v4}
{DeepSeek-AI}.
\newblock Deepseek-v4: Towards highly efficient million-token context
  intelligence.
\newblock Technical report, 2026.
\newblock Available at Hugging Face model repository. Accessed: 2026-06-10.

\bibitem{ding2026fmagent}
Haoran Ding, Zhaoguo Wang, and Haibo Chen.
\newblock Fm-agent: Scaling formal methods to large systems via llm-based
  hoare-style reasoning, 2026.

\bibitem{adaptcache}
Shaoting Feng, Hanchen Li, Kuntai Du, Zhuohan Gu, Yuhan Liu, Jiayi Yao,
  Siddhant Ray, Samuel Shen, Yihua Cheng, Ganesh Ananthanarayanan, and Junchen
  Jiang.
\newblock Adaptcache: {KV} cache native storage hierarchy for low-delay and
  high-quality language model serving.
\newblock {\em CoRR}, abs/2509.00105, 2025.

\bibitem{DBLP:conf/usenix/GaoHSKJDYYZ24}
Bin Gao, Zhuomin He, Puru Sharma, Qingxuan Kang, Djordje Jevdjic, Junbo Deng,
  Xingkun Yang, Zhou Yu, and Pengfei Zuo.
\newblock Cost-efficient large language model serving for multi-turn
  conversations with cachedattention.
\newblock In Saurabh Bagchi and Yiying Zhang, editors, {\em Proceedings of the
  2024 {USENIX} Annual Technical Conference, {USENIX} {ATC} 2024, Santa Clara,
  CA, USA, July 10-12, 2024}, pages 111--126. {USENIX} Association, 2024.

\bibitem{DBLP:conf/mlsys/GimCLSK024}
In~Gim, Guojun Chen, Seung{-}Seob Lee, Nikhil Sarda, Anurag Khandelwal, and Lin
  Zhong.
\newblock Prompt cache: Modular attention reuse for low-latency inference.
\newblock In Phillip~B. Gibbons, Gennady Pekhimenko, and Christopher~De Sa,
  editors, {\em Proceedings of the Seventh Annual Conference on Machine
  Learning and Systems, MLSys 2024, Santa Clara, CA, USA, May 13-16, 2024}.
  mlsys.org, 2024.

\bibitem{DBLP:journals/corr/abs-2602-00328}
Nikhil Gopal and Kostis Kaffes.
\newblock Harvest: Opportunistic peer-to-peer {GPU} caching for {LLM}
  inference.
\newblock {\em CoRR}, abs/2602.00328, 2026.

\bibitem{DBLP:conf/osdi/GujaratiKAHKVM20}
Arpan Gujarati, Reza Karimi, Safya Alzayat, Wei Hao, Antoine Kaufmann, Ymir
  Vigfusson, and Jonathan Mace.
\newblock Serving dnns like clockwork: Performance predictability from the
  bottom up.
\newblock In {\em 14th {USENIX} Symposium on Operating Systems Design and
  Implementation, {OSDI} 2020, Virtual Event, November 4-6, 2020}, pages
  443--462. {USENIX} Association, 2020.

\bibitem{DBLP:conf/nips/HooperKMMSKG24}
Coleman Hooper, Sehoon Kim, Hiva Mohammadzadeh, Michael~W. Mahoney,
  Yakun~Sophia Shao, Kurt Keutzer, and Amir Gholami.
\newblock Kvquant: Towards 10 million context length {LLM} inference with {KV}
  cache quantization.
\newblock In Amir Globersons, Lester Mackey, Danielle Belgrave, Angela Fan,
  Ulrich Paquet, Jakub~M. Tomczak, and Cheng Zhang, editors, {\em Advances in
  Neural Information Processing Systems 37: Annual Conference on Neural
  Information Processing Systems 2024, NeurIPS 2024, Vancouver, BC, Canada,
  December 10 - 15, 2024}, 2024.

\bibitem{ascendra}
Azam Ikram, Xiang Li, Sameh Elnikety, and Saurabh Bagchi.
\newblock Ascendra: Dynamic request prioritization for efficient {LLM} serving.
\newblock {\em CoRR}, abs/2504.20828, 2025.

\bibitem{codex-swebenchpro-trace}
Inferact.
\newblock Codex agent traces on {SWE-Bench Pro}.
\newblock Hugging Face dataset,
  \url{https://huggingface.co/datasets/Inferact/codex_swebenchpro_traces},
  2026.

\bibitem{DBLP:conf/asplos/KamathPM0RP25}
Aditya~K. Kamath, Ramya Prabhu, Jayashree Mohan, Simon Peter, Ramachandran
  Ramjee, and Ashish Panwar.
\newblock Pod-attention: Unlocking full prefill-decode overlap for faster {LLM}
  inference.
\newblock In Lieven Eeckhout, Georgios Smaragdakis, Katai Liang, Adrian
  Sampson, Martha~A. Kim, and Christopher~J. Rossbach, editors, {\em
  Proceedings of the 30th {ACM} International Conference on Architectural
  Support for Programming Languages and Operating Systems, Volume 2, {ASPLOS}
  2025, Rotterdam, Netherlands, 30 March 2025 - 3 April 2025}, pages 897--912.
  {ACM}, 2025.

\bibitem{DBLP:conf/mlsys/KangBHKRR25}
Hao Kang, Srikant Bharadwaj, James Hensman, Tushar Krishna, Victor R{\"{u}}hle,
  and Saravan Rajmohan.
\newblock Turboattention: Efficient attention approximation for high
  throughputs llm.
\newblock In Matei Zaharia, Gauri Joshi, and Yingyan~(Celine) Lin, editors,
  {\em Proceedings of the Eighth Conference on Machine Learning and Systems,
  MLSys 2025, Santa Clara, CA, USA, May 12-15, 2025}. OpenReview.net/mlsys.org,
  2025.

\bibitem{aqua}
Abhishek~Vijaya Kumar, Gianni Antichi, and Rachee Singh.
\newblock Aqua: Network-accelerated memory offloading for {LLMs} in scale-up
  {GPU} domains.
\newblock In {\em Proceedings of the 30th {ACM} International Conference on
  Architectural Support for Programming Languages and Operating Systems, Volume
  2, {ASPLOS} 2025, Rotterdam, Netherlands, 30 March 2025 - 3 April 2025},
  pages 48--62. {ACM}, 2025.

\bibitem{vllm}
Woosuk Kwon, Zhuohan Li, Siyuan Zhuang, Ying Sheng, Lianmin Zheng, Cody~Hao Yu,
  Joseph Gonzalez, Hao Zhang, and Ion Stoica.
\newblock Efficient memory management for large language model serving with
  pagedattention.
\newblock In {\em Proceedings of the 29th Symposium on Operating Systems
  Principles, {SOSP} 2023, Koblenz, Germany, October 23-26, 2023}, pages
  611--626. {ACM}, 2023.

\bibitem{DBLP:journals/corr/abs-2507-11507}
Ruihao Li, Shagnik Pal, Vineeth~Narayan Pullu, Prasoon Sinha, Jeeho Ryoo,
  Lizy~K. John, and Neeraja~J. Yadwadkar.
\newblock {MIRAGE:} {KV} cache optimization through parameter remapping for
  multi-tenant {LLM} serving.
\newblock {\em CoRR}, abs/2507.11507, 2025.

\bibitem{DBLP:conf/nips/LiHYVLYCLC24}
Yuhong Li, Yingbing Huang, Bowen Yang, Bharat Venkitesh, Acyr Locatelli,
  Hanchen Ye, Tianle Cai, Patrick Lewis, and Deming Chen.
\newblock Snapkv: {LLM} knows what you are looking for before generation.
\newblock In Amir Globersons, Lester Mackey, Danielle Belgrave, Angela Fan,
  Ulrich Paquet, Jakub~M. Tomczak, and Cheng Zhang, editors, {\em Advances in
  Neural Information Processing Systems 37: Annual Conference on Neural
  Information Processing Systems 2024, NeurIPS 2024, Vancouver, BC, Canada,
  December 10 - 15, 2024}, 2024.

\bibitem{adaserve}
Zikun Li, Zhuofu Chen, Remi Delacourt, Gabriele Oliaro, Zeyu Wang, Qinghan
  Chen, Shuhuai Lin, April Yang, Zhihao Zhang, Zhuoming Chen, Yi-Hsiang Lai,
  Xinhao Cheng, Xupeng Miao, and Zhihao Jia.
\newblock Adaserve: Accelerating multi-{SLO} {LLM} serving with
  {SLO}-customized speculative decoding.
\newblock In {\em Proceedings of the 21st European Conference on Computer
  Systems, EuroSys 2026, McEwan Hall/The University of Edinburgh, Edinburgh,
  Scotland, UK, April 27-30, 2026}, pages 36--54. {ACM}, 2026.

\bibitem{DBLP:journals/corr/abs-2501-13111}
Dimitrios Liakopoulos, Tianrui Hu, Prasoon Sinha, and Neeraja~J. Yadwadkar.
\newblock iserve: An intent-based serving system for llms.
\newblock {\em CoRR}, abs/2501.13111, 2025.

\bibitem{DBLP:conf/osdi/LinHZ00CQ24}
Chaofan Lin, Zhenhua Han, Chengruidong Zhang, Yuqing Yang, Fan Yang, Chen Chen,
  and Lili Qiu.
\newblock Parrot: Efficient serving of llm-based applications with semantic
  variable.
\newblock In Ada Gavrilovska and Douglas~B. Terry, editors, {\em 18th {USENIX}
  Symposium on Operating Systems Design and Implementation, {OSDI} 2024, Santa
  Clara, CA, USA, July 10-12, 2024}, pages 929--945. {USENIX} Association,
  2024.

\bibitem{andes}
Jiachen Liu, Zhiyu Wu, Jae-Won Chung, Fan Lai, Myungjin Lee, and Mosharaf
  Chowdhury.
\newblock Andes: Defining and enhancing quality-of-experience in {LLM}-based
  text streaming services.
\newblock {\em CoRR}, abs/2404.16283, 2024.

\bibitem{DBLP:conf/sigcomm/LiuLCRHZDY0AMHH24}
Yuhan Liu, Hanchen Li, Yihua Cheng, Siddhant Ray, Yuyang Huang, Qizheng Zhang,
  Kuntai Du, Jiayi Yao, Shan Lu, Ganesh Ananthanarayanan, Michael Maire, Henry
  Hoffmann, Ari Holtzman, and Junchen Jiang.
\newblock Cachegen: {KV} cache compression and streaming for fast large
  language model serving.
\newblock In {\em Proceedings of the {ACM} {SIGCOMM} 2024 Conference, {ACM}
  {SIGCOMM} 2024, Sydney, NSW, Australia, August 4-8, 2024}, pages 38--56.
  {ACM}, 2024.

\bibitem{DBLP:conf/icml/LiuYJZXBC024}
Zirui Liu, Jiayi Yuan, Hongye Jin, Shaochen~(Henry) Zhong, Zhaozhuo Xu,
  Vladimir Braverman, Beidi Chen, and Xia Hu.
\newblock {KIVI:} {A} tuning-free asymmetric 2bit quantization for {KV} cache.
\newblock In Ruslan Salakhutdinov, Zico Kolter, Katherine~A. Heller, Adrian
  Weller, Nuria Oliver, Jonathan Scarlett, and Felix Berkenkamp, editors, {\em
  Forty-first International Conference on Machine Learning, {ICML} 2024,
  Vienna, Austria, July 21-27, 2024}, Proceedings of Machine Learning Research,
  pages 32332--32344. {PMLR} / OpenReview.net, 2024.

\bibitem{llm-d}
llm-d.
\newblock \burl{https://github.com/llm-d/llm-d}, 2026.

\bibitem{lmcache}
Lmcache: The best kv cache layer for enterprise-scale llm inference.
\newblock \burl{https://github.com/LMCache/LMCache}, 2026.

\bibitem{DBLP:journals/corr/abs-2502-13965}
Michael Luo, Xiaoxiang Shi, Colin Cai, Tianjun Zhang, Justin Wong, Yichuan
  Wang, Chi Wang, Yanping Huang, Zhifeng Chen, Joseph~E. Gonzalez, and Ion
  Stoica.
\newblock Autellix: An efficient serving engine for {LLM} agents as general
  programs.
\newblock {\em CoRR}, abs/2502.13965, 2025.

\bibitem{helix}
Yixuan Mei, Yonghao Zhuang, Xupeng Miao, Juncheng Yang, Zhihao Jia, and Rashmi
  Vinayak.
\newblock Helix: Serving large language models over heterogeneous {GPUs} and
  network via max-flow.
\newblock In {\em Proceedings of the 30th {ACM} International Conference on
  Architectural Support for Programming Languages and Operating Systems, Volume
  1, {ASPLOS} 2025, Rotterdam, The Netherlands, 30 March 2025 - 3 April 2025},
  pages 586--602. {ACM}, 2025.

\bibitem{DBLP:conf/asplos/MiaoSDXL0J24}
Xupeng Miao, Chunan Shi, Jiangfei Duan, Xiaoli Xi, Dahua Lin, Bin Cui, and
  Zhihao Jia.
\newblock Spotserve: Serving generative large language models on preemptible
  instances.
\newblock In Rajiv Gupta, Nael~B. Abu{-}Ghazaleh, Madan Musuvathi, and Dan
  Tsafrir, editors, {\em Proceedings of the 29th {ACM} International Conference
  on Architectural Support for Programming Languages and Operating Systems,
  Volume 2, {ASPLOS} 2024, La Jolla, CA, USA, 27 April 2024- 1 May 2024}, pages
  1112--1127. {ACM}, 2024.

\bibitem{minimax-m3}
{MiniMax}.
\newblock {MiniMax-M3}.
\newblock \url{https://www.minimax.io/blog/minimax-m3}, 2026.

\bibitem{power-of-two}
Michael Mitzenmacher.
\newblock The power of two choices in randomized load balancing.
\newblock {\em {IEEE} Trans. Parallel Distributed Syst.}, 12(10):1094--1104,
  2001.

\bibitem{kimi2026agentswarm}
{Moonshot AI}.
\newblock Kimi introduces agent swarm: Let 100 {AI} agents work for you.
\newblock \url{https://www.kimi.com/blog/agent-swarm}, 2026.

\bibitem{ai-Dynamo}
NVIDIA.
\newblock ai-dynamo.
\newblock \burl{https://github.com/ai-dynamo/dynamo}, 2026.

\bibitem{openai-api}
OpenAI.
\newblock {OpenAI} {API}.
\newblock \burl{https://openai.com/api/}, 2026.

\bibitem{openai-responses}
OpenAI.
\newblock {OpenAI} {API} docs: Conversation state.
\newblock
  \url{https://developers.openai.com/api/docs/guides/conversation-state}, 2026.

\bibitem{openclaw}
{OpenClaw}.
\newblock Openclaw.
\newblock \url{https://openclaw.ai}, 2026.

\bibitem{unicache}
Bei Ouyang, Yifan Qiao, and Jiarong Xing.
\newblock Unicache: Unifying prefix cache eviction for heterogeneous {LLM}
  serving workloads.
\newblock {\em Proc. {ACM} Meas. Anal. Comput. Syst.}, 10(2):54:1--54:27, 2026.

\bibitem{DBLP:conf/isca/PatelCZSGMB24}
Pratyush Patel, Esha Choukse, Chaojie Zhang, Aashaka Shah, {\'{I}}{\~{n}}igo
  Goiri, Saeed Maleki, and Ricardo Bianchini.
\newblock Splitwise: Efficient generative {LLM} inference using phase
  splitting.
\newblock In {\em 51st {ACM/IEEE} Annual International Symposium on Computer
  Architecture, {ISCA} 2024, Buenos Aires, Argentina, June 29 - July 3, 2024},
  pages 118--132. {IEEE}, 2024.

\bibitem{DBLP:conf/ics/PearceGSSA12}
Olga Pearce, Todd Gamblin, Bronis~R. de~Supinski, Martin Schulz, and Nancy~M.
  Amato.
\newblock Quantifying the effectiveness of load balance algorithms.
\newblock In Utpal Banerjee, Kyle~A. Gallivan, Gianfranco Bilardi, and Manolis
  Katevenis, editors, {\em International Conference on Supercomputing, ICS'12,
  Venice, Italy, June 25-29, 2012}, pages 185--194. {ACM}, 2012.

\bibitem{DBLP:conf/asplos/PrabhuNMRP25}
Ramya Prabhu, Ajay Nayak, Jayashree Mohan, Ramachandran Ramjee, and Ashish
  Panwar.
\newblock vattention: Dynamic memory management for serving llms without
  pagedattention.
\newblock In Lieven Eeckhout, Georgios Smaragdakis, Kaitai Liang, Adrian
  Sampson, Martha~A. Kim, and Christopher~J. Rossbach, editors, {\em
  Proceedings of the 30th {ACM} International Conference on Architectural
  Support for Programming Languages and Operating Systems, Volume 1, {ASPLOS}
  2025, Rotterdam, The Netherlands, 30 March 2025 - 3 April 2025}, pages
  1133--1150. {ACM}, 2025.

\bibitem{305212}
Ruoyu Qin, Zheming Li, Weiran He, Jialei Cui, Feng Ren, Mingxing Zhang, Yongwei
  Wu, Weimin Zheng, and Xinran Xu.
\newblock Mooncake: Trading more storage for less computation {\textemdash} a
  {KVCache-centric} architecture for serving {LLM} chatbot.
\newblock In {\em 23rd USENIX Conference on File and Storage Technologies (FAST
  25)}, pages 155--170, Santa Clara, CA, February 2025. USENIX Association.

\bibitem{qwen3.7}
{Qwen Team}.
\newblock Qwen3.7.
\newblock \url{https://qwen.ai/blog?id=qwen3.7}, 2026.

\bibitem{DBLP:journals/corr/abs-2504-09285}
Chaoyi Ruan, Yinhe Chen, Dongqi Tian, Yandong Shi, Yongji Wu, Jialin Li, and
  Cheng Li.
\newblock Dynaserve: Unified and elastic tandem-style execution for dynamic
  disaggregated {LLM} serving.
\newblock {\em CoRR}, abs/2504.09285, 2025.

\bibitem{traceupscaler}
Sultan~Mahmud Sajal, Timothy Zhu, Bhuvan Urgaonkar, and Siddhartha Sen.
\newblock Traceupscaler: Upscaling traces to evaluate systems at high load.
\newblock In {\em Proceedings of the Nineteenth European Conference on Computer
  Systems, EuroSys 2024, Athens, Greece, April 22-25, 2024}, pages 942--961.
  {ACM}, 2024.

\bibitem{dontstopmenow}
Rana Shahout, Eran Malach, Chunwei Liu, Weifan Jiang, Minlan Yu, and Michael
  Mitzenmacher.
\newblock Don't stop me now: Embedding based scheduling for {LLMs}.
\newblock In {\em The Thirteenth International Conference on Learning
  Representations, {ICLR} 2025, Singapore, April 24-28, 2025}. OpenReview.net,
  2025.

\bibitem{vtc}
Ying Sheng, Shiyi Cao, Dacheng Li, Banghua Zhu, Zhuohan Li, Danyang Zhuo,
  Joseph~E. Gonzalez, and Ion Stoica.
\newblock Fairness in serving large language models.
\newblock In {\em 18th {USENIX} Symposium on Operating Systems Design and
  Implementation, {OSDI} 2024, Santa Clara, CA, USA, July 10-12, 2024}, pages
  965--988. {USENIX} Association, 2024.

\bibitem{DBLP:conf/iclr/SrivatsaHAL025}
Vikranth Srivatsa, Zijian He, Reyna Abhyankar, Dongming Li, and Yiying Zhang.
\newblock Preble: Efficient distributed prompt scheduling for {LLM} serving.
\newblock In {\em The Thirteenth International Conference on Learning
  Representations, {ICLR} 2025, Singapore, April 24-28, 2025}. OpenReview.net,
  2025.

\bibitem{DBLP:conf/osdi/SunHZXZL024}
Biao Sun, Ziming Huang, Hanyu Zhao, Wencong Xiao, Xinyi Zhang, Yong Li, and Wei
  Lin.
\newblock Llumnix: Dynamic scheduling for large language model serving.
\newblock In Ada Gavrilovska and Douglas~B. Terry, editors, {\em 18th {USENIX}
  Symposium on Operating Systems Design and Implementation, {OSDI} 2024, Santa
  Clara, CA, USA, July 10-12, 2024}, pages 173--191. {USENIX} Association,
  2024.

\bibitem{vllm-code}
vllm v0.12.0 release.
\newblock \burl{https://github.com/vllm-project/vllm/releases/tag/v0.12.0},
  2025.

\bibitem{taichi}
Chao Wang, Pengfei Zuo, Zhangyu Chen, Yunkai Liang, Zhou Yu, and Ming-Chang
  Yang.
\newblock Prefill-decode aggregation or disaggregation? unifying both for
  goodput-optimized {LLM} serving.
\newblock {\em CoRR}, abs/2508.01989, 2025.

\bibitem{kvcache}
Jiahao Wang, Jinbo Han, Xingda Wei, Sijie Shen, Dingyan Zhang, Chenguang Fang,
  Rong Chen, Wenyuan Yu, and Haibo Chen.
\newblock Kvcache cache in the wild: Characterizing and optimizing kvcache
  cache at a large cloud provider.
\newblock In {\em 2025 USENIX Annual Technical Conference (USENIX ATC 25)}.
  USENIX Association, July 2025.

\bibitem{DBLP:conf/sosp/WuLZ0L024}
Bingyang Wu, Shengyu Liu, Yinmin Zhong, Peng Sun, Xuanzhe Liu, and Xin Jin.
\newblock Loongserve: Efficiently serving long-context large language models
  with elastic sequence parallelism.
\newblock In Emmett Witchel, Christopher~J. Rossbach, Andrea~C.
  Arpaci{-}Dusseau, and Kimberly Keeton, editors, {\em Proceedings of the {ACM}
  {SIGOPS} 30th Symposium on Operating Systems Principles, {SOSP} 2024, Austin,
  TX, USA, November 4-6, 2024}, pages 640--654. {ACM}, 2024.

\bibitem{skywalker}
Tian Xia, Ziming Mao, Jamison Kerney, Ethan~J. Jackson, Zhifei Li, Jiarong
  Xing, Scott Shenker, and Ion Stoica.
\newblock Skywalker: A locality-aware cross-region load balancer for {LLM}
  inference.
\newblock In {\em Proceedings of the 21st European Conference on Computer
  Systems, EuroSys 2026, McEwan Hall/The University of Edinburgh, Edinburgh,
  Scotland, UK, April 27-30, 2026}, pages 1349--1364. {ACM}, 2026.

\bibitem{shadowserve}
Xingyu Xiang, Raj Joshi, Yuhan Liu, Jiayi Yao, Chenxingyu Zhao, Junchen Jiang,
  Yang Zhou, Eddie Kohler, and Minlan Yu.
\newblock Shadowserve: Interference-free {KV} cache fetching for distributed
  prefix caching.
\newblock {\em CoRR}, abs/2509.16857, 2025.

\bibitem{DBLP:conf/sosp/Xiang0QYZYZL0025}
Yuxing Xiang, Xue Li, Kun Qian, Yufan Yang, Diwen Zhu, Wenyuan Yu, Ennan Zhai,
  Xuanzhe Liu, Xin Jin, and Jingren Zhou.
\newblock Aegaeon: Effective {GPU} pooling for concurrent {LLM} serving on the
  market.
\newblock In Youjip Won, Youngjin Kwon, Ding Yuan, and Rebecca Isaacs, editors,
  {\em Proceedings of the {ACM} {SIGOPS} 31st Symposium on Operating Systems
  Principles, {SOSP} 2025, Lotte Hotel World, Seoul, Republic of Korea, October
  13-16, 2025}, pages 1030--1045. {ACM}, 2025.

\bibitem{servegen}
Yuxing Xiang, Xue Li, Kun Qian, Yan Zhang, Wenyuan Yu, Ennan Zhai, Xin Jin, and
  Jingren Zhou.
\newblock Servegen: Workload characterization and generation of large language
  model serving in production.
\newblock In {\em 23rd {USENIX} Symposium on Networked Systems Design and
  Implementation, {NSDI} 2026, Renton, WA, May 4-6, 2026}, pages 1845--1859.
  {USENIX} Association, 2026.

\bibitem{CacheBlend}
Jiayi Yao, Hanchen Li, Yuhan Liu, Siddhant Ray, Yihua Cheng, Qizheng Zhang,
  Kuntai Du, Shan Lu, and Junchen Jiang.
\newblock Cacheblend: Fast large language model serving for {RAG} with cached
  knowledge fusion.
\newblock In {\em Proceedings of the Twentieth European Conference on Computer
  Systems, EuroSys 2025, Rotterdam, The Netherlands, 30 March 2025 - 3 April
  2025}, pages 94--109. {ACM}, 2025.

\bibitem{DBLP:conf/acl/YeTHL24}
Lu~Ye, Ze~Tao, Yong Huang, and Yang Li.
\newblock Chunkattention: Efficient self-attention with prefix-aware {KV} cache
  and two-phase partition.
\newblock In Lun{-}Wei Ku, Andre Martins, and Vivek Srikumar, editors, {\em
  Proceedings of the 62nd Annual Meeting of the Association for Computational
  Linguistics (Volume 1: Long Papers), {ACL} 2024, Bangkok, Thailand, August
  11-16, 2024}, pages 11608--11620. Association for Computational Linguistics,
  2024.

\bibitem{DBLP:conf/mlsys/00010LLZW0KGKC25}
Zihao Ye, Lequn Chen, Ruihang Lai, Wuwei Lin, Yineng Zhang, Stephanie Wang,
  Tianqi Chen, Baris Kasikci, Vinod Grover, Arvind Krishnamurthy, and Luis
  Ceze.
\newblock Flashinfer: Efficient and customizable attention engine for {LLM}
  inference serving.
\newblock In Matei Zaharia, Gauri Joshi, and Yingyan~(Celine) Lin, editors,
  {\em Proceedings of the Eighth Conference on Machine Learning and Systems,
  MLSys 2025, Santa Clara, CA, USA, May 12-15, 2025}. OpenReview.net/mlsys.org,
  2025.

\bibitem{DBLP:journals/corr/abs-2312-05516}
Lingfan Yu and Jinyang Li.
\newblock Stateful large language model serving with pensieve.
\newblock {\em CoRR}, abs/2312.05516, 2023.

\bibitem{pythia}
Shan Yu, Junyi Shu, Yuanjiang Ni, Kun Qian, Xue Li, Yang Wang, Jinyuan Zhang,
  Ziyi Xu, Shuo Yang, Lingjun Zhu, Ennan Zhai, Qingda Lu, Jiarong Xing, Youyou
  Lu, Xin Jin, Xuanzhe Liu, and Harry Xu.
\newblock Pythia: Exploiting workflow predictability for efficient agent-native
  {LLM} serving.
\newblock {\em CoRR}, abs/2604.25899, 2026.

\bibitem{DBLP:journals/corr/abs-2602-06502}
Ying Yuan, Pengfei Zuo, Bo~Wang, Zhangyu Chen, Zhipeng Tan, and Zhou Yu.
\newblock Dualmap: Enabling both cache affinity and load balancing for
  distributed {LLM} serving.
\newblock {\em CoRR}, abs/2602.06502, 2026.

\bibitem{zhang2026simple}
Dingyan Zhang, Jinbo Han, Kaixi Zhang, Xingda Wei, Sijie Shen, Chenguang Fang,
  Wenyuan Yu, Jingren Zhou, and Rong Chen.
\newblock Simple is better: Multiplication may be all you need for llm request
  scheduling.
\newblock In {\em Proceedings of the 20th USENIX Symposium on Operating Systems
  Design and Implementation (OSDI '26)}, Seattle, WA, USA, July 2026. USENIX
  Association.

\bibitem{DBLP:conf/osdi/ZhangWLWS0025}
Dingyan Zhang, Haotian Wang, Yang Liu, Xingda Wei, Yizhou Shan, Rong Chen, and
  Haibo Chen.
\newblock Blitzscale: Fast and live large model autoscaling with {O(1)} host
  caching.
\newblock In Lidong Zhou and Yuanyuan Zhou, editors, {\em 19th {USENIX}
  Symposium on Operating Systems Design and Implementation, {OSDI} 2025,
  Boston, MA, USA, July 7-9, 2025}, pages 275--293. {USENIX} Association, 2025.

\bibitem{agentkv}
Rui Zhang, Chaeeun Kim, Shaoting Feng, Kuntai Du, Yuhan Liu, Yi~Zhong,
  Cheng-Wei Ching, Junchen Jiang, and Liting Hu.
\newblock Learning agent execution for {KV}-cache management in agentic
  serving.
\newblock {\em CoRR}, abs/2608.14624, 2026.

\bibitem{DBLP:journals/corr/abs-2412-03131}
Yanqi Zhang, Yuwei Hu, Runyuan Zhao, John C.~S. Lui, and Haibo Chen.
\newblock Unifying {KV} cache compression for large language models with
  leankv.
\newblock {\em CoRR}, abs/2412.03131, 2024.

\bibitem{sglang}
Lianmin Zheng, Liangsheng Yin, Zhiqiang Xie, Jeff Huang, Chuyue Sun, Cody~Hao
  Yu, Shiyi Cao, Christos Kozyrakis, Ion Stoica, Joseph~E. Gonzalez, Clark~W.
  Barrett, and Ying Sheng.
\newblock Efficiently programming large language models using sglang.
\newblock {\em CoRR}, abs/2312.07104, 2023.

\bibitem{DBLP:conf/osdi/ZhongLCHZL0024}
Yinmin Zhong, Shengyu Liu, Junda Chen, Jianbo Hu, Yibo Zhu, Xuanzhe Liu, Xin
  Jin, and Hao Zhang.
\newblock Distserve: Disaggregating prefill and decoding for goodput-optimized
  large language model serving.
\newblock In {\em 18th {USENIX} Symposium on Operating Systems Design and
  Implementation, {OSDI} 2024, Santa Clara, CA, USA, July 10-12, 2024}, pages
  193--210. {USENIX} Association, 2024.

\bibitem{DBLP:conf/osdi/ZhuGZZZGX0KLWWK25}
Kan Zhu, Yufei Gao, Yilong Zhao, Liangyu Zhao, Gefei Zuo, Yile Gu, Dedong Xie,
  Zihao Ye, Keisuke Kamahori, Chien{-}Yu Lin, Ziren Wang, Stephanie Wang,
  Arvind Krishnamurthy, and Baris Kasikci.
\newblock Nanoflow: Towards optimal large language model serving throughput.
\newblock In Lidong Zhou and Yuanyuan Zhou, editors, {\em 19th {USENIX}
  Symposium on Operating Systems Design and Implementation, {OSDI} 2025,
  Boston, MA, USA, July 7-9, 2025}, pages 749--765. {USENIX} Association, 2025.

\bibitem{DBLP:journals/corr/abs-2507-17769}
Kan Zhu, Haiyang Shi, Le~Xu, Jiaxin Shan, Arvind Krishnamurthy, Baris Kasikci,
  and Liguang Xie.
\newblock Polyserve: Efficient multi-slo serving at scale.
\newblock {\em CoRR}, abs/2507.17769, 2025.

\end{thebibliography}
